\documentclass[twocolumn]{aastex7}

\usepackage{amsmath} 
\usepackage{subfigure}
\begin{document}

\title{Subparsec Jet Morphologies of M87 at 43 GHz: Effects of Asymmetric Plasmoids within the Jet}

\author[orcid=0000-0002-3477-5184,sname='Li']{Guan-Hong Li}
\affiliation{Department of Physics, National Taiwan Normal University, No. 88, Sec. 4, Tingzhou Rd., Taipei 116059, Taiwan, R.O.C.}
\affiliation{Graduate Institute of Astrophysics, National Taiwan University, No. 1, Sec. 4, Roosevelt Rd., Taipei 106216, Taiwan, R.O.C.}
\email{ghli@gapps.ntnu.edu.tw}

\author[orcid=0000-0001-9270-8812,sname='Pu']{Hung-Yi Pu}
\affiliation{Department of Physics, National Taiwan Normal University, No. 88, Sec. 4, Tingzhou Rd., Taipei 116059, Taiwan, R.O.C.}
\affiliation{Center of Astronomy and Gravitation, National Taiwan Normal University, No. 88, Sec. 4, Tingzhou Road, Taipei 116059, Taiwan, R.O.C.}
\affiliation{Institute of Astronomy and Astrophysics, Academia Sinica, 11F of Astronomy-Mathematics Building, AS/NTU No. 1, Sec. 4, Roosevelt Rd., Taipei 106216, Taiwan, R.O.C.}
\email{hypu@gapps.ntnu.edu.tw}

\begin{abstract}
Radio observations of the M87 jet reveal limb-brightened features that exhibit temporal variations, with the brighter limb side not remaining consistently fixed at the subparsec scales. Utilizing a force-free jet model that considers exclusively the relativistic plasma dynamics along large-scale magnetic fields attached onto the central black hole, we examine the effects of asymmetric plasmoid injections within the subparsec jet. At subparsec scales, the jet velocity is predominantly influenced by poloidal velocity, leading to distinctive characteristics in both the morphology and trajectories of the plasmoids injected within the jet.
We explore the potential modifications to the limb-brightened subparsec jet images resulting from the injection of asymmetric, shearing plasma, which is traditionally considered solely under conditions of stationary and symmetric mass loading. By comparing the model jet properties with the 43 GHz observations of the M87 jet, it is suggested that the asymmetric injection of plasmoids within the jet offers a satisfactory explanation for the observed velocity variations and morphological dynamics of the M87 jet.

\end{abstract}


\section{Introduction}
Resolved by the very long baseline interferometry  observations, the prominent relativistic jet launched by the supermassive black hole at the center of the giant elliptical galaxy Messier 87 (M87) yields crucial insights into the mechanisms responsible for launching low-luminosity active galactic nuclei (LLAGNs).
These observations encompass phenomena such as jet morphology \citep[e.g.][]{kovalev2007,hada2016high,walker2018structure,kim2018,lu2023}, jet width and collimation \citep[e.g.][]{asada2012,hada2013innermost}, jet acceleration and velocity \citep[e.g.][]{mer2016,nakamura2018,park2019}. 
Located at a distance of 16.8 Mpc, the M87 black hole has a mass of $M_{\rm BH}\simeq   6.5\times 10^{9}M_\odot$ \citep[e.g.][]{eht2019a}, which corresponds to a scale $1~{\rm mas}\approx0.08~{\rm pc}\approx262~R_{\rm g}$, where $R_{\rm g}\equiv GM_{\rm BH}/c^{2}$.

The M87 jet exhibits limb-brightened features with time-dependent behaviors.
\citet[][]{walker2018structure} presented monitoring observations of M87 with Very Long Baseline Array at 43 GHz with a resolution of about 100 $R_{\rm g}$ and revealed significant outward motions of blobs in the jet \citep[see also][]{walker2008_movie}.
Interestingly, while the southern limb of the M87 jet is usually brighter at the parsec scale, the northern limb may appear brighter sometimes \citep[see the online version of Figure 9 and 11 of][for a movie of the jet activities]{walker2018structure}. 
The visual inspection of the kinetics of the subparsc jet of M87 by \citet{walker2018structure} has been analyzed by \citet{mer2016} with the wavelet-based image segmentation and evaluation (WISE) method, which reveals velocity patterns dominated by poloidal motions. 

The observed phenomenon of a typically brighter southern limb of the M87 jet near the core aligns with the geometric configuration of the M87 black hole system. Based on the position angle of the right-like structure at the horizon scale, \citet[][]{2019ehtV} interprets that the rotational axis of the M87 black hole is directed away from Earth. Assuming the jet axis is approximately parallel to the rotational axis of the black hole, the toroidal motion of the plasma within the jet would inherently amplify the approaching side of the jet, thereby causing the southern limb to appear brighter than the northern limb, as demonstrated in numerous studies of model M87 jet images by postprocessing radiative mechanisms from numerical simulations of M87 black hole systems \citep[e.g.][]{monika2017,chael2019,jordy2019,cruz2022,fromm2022}.

In the theoretical modeling of the synchrotron jet emission from LLAGNs, the loading of energetic nonthermal electrons within the jet represents one of the most uncertain factors \citep[see, e.g.,][for a review]{pu2022modeling}.  
Consequently, an explanation for the observed variability in the brightness of the jet limbs and the dynamical evolution of jet images may lie in the time-dependent asymmetric injection of plasmoids within the jet. However, details of the formation mechanism and subsequent evolution of these plasmoids, such as the size of the injected location, the injection cadence, and the afterward shearing and cooling effects, are not sufficiently understood. As a result, it is still challenging to properly include plasmoid injection physics in numerical simulations.

In this work, we aim to investigate the influence of asymmetric plasmoids injected within the M87 jet, taking into account how the properties of the injected plasmoid, such as their motion and their shape due to the shearing effect, would be associated with the dynamics of the jet.
We are especially interested in exploring whether the observed dynamical jet morphology and the trajectories within the M87 jet at 43 GHz \citep{walker2018structure,mer2016} can be reproduced by combining existing well-established jet models with possible injections of plasmoids.
In our working proposal, to examine the emission characteristics of relativistic jets at scales significantly exceeding the event horizon, gravitational effects are neglected. The axisymmetric jet structure is modeled using the semianalytical force-free jet model within Minkowski spacetime, as detailed in \citet{tchekhovskoy2008simulations}, \cite{broderick2009imaging} and \cite{takahashi2018fast}, with subsequent modifications.
Using the constructed magnetic and velocity fields, we first build a synchrotron image that closely approximates the observed image. 
The influence of injected, shearing plasmoids is then modeled and added onto such a background to examine the diversity of the resulting jet morphologies and the alterations to the overall flux. 
In addition, we simulate the trajectories of the injected plasmoids within the jet.

The structure of this paper is as follows. Section \S \ref{sec:jet_model} presents the jet model along with the details of our model setup. Section \S \ref{sec:comparion} discusses how the jet morphology varies with the injected plasmoid within the jet, the apparent motion of the plasmoid, and comparisons with observational data. Finally, a summary and implications are presented in Section \S \ref{sec:summary}.

\section{Modeling M87 jet}\label{sec:jet_model}
Our jet model is based on the force-free jet model \citep{tchekhovskoy2008simulations,broderick2009imaging,takahashi2018fast}, with further modifications. 
 For completeness, we first summarize the force-free jet model \citep{tchekhovskoy2008simulations,broderick2009imaging,takahashi2018fast} in \S \ref{sec:ffjet}. Following the introduction of mass loading in \S \ref{sec:mass_loading}, the modifications and specifics of our model setup are provided in \S \ref{sec:model_setup}. Throughout the paper, we set $c=1$.

\subsection{Force-free Jet Structure}\label{sec:ffjet}
To characterize the jet structure, we start with the approximated solution of the force-free Grad-Shafranov equation in spherical coordinates $(r,\theta,\phi)$\citep{narayan2007,tchekhovskoy2008simulations} 
\begin{equation}\label{eq:streamline}
    \Psi = r^{\nu}(1-\cos\theta)\;.
\end{equation}
The aforementioned streamline function describes large-scale magnetic fields, which constitute the jet. 
The parameter $\nu$ is associated with the jet configuration: $\nu = 0$ results in a split monopole-like magnetic field, while $\nu = 1$ leads to a parabolic magnetic field. 

\subsubsection{Magnetic Field Structure}
The magnetic field is related to the stream function by
\begin{equation}\label{eq:bphi}
    B_\phi = -\frac{2\Omega_F\Psi}{R}
\end{equation}
\begin{equation}\label{eq:br}
    B_r = r^{\nu-2}\;,
\end{equation}
\begin{equation}\label{eq:bth}
    B_\theta = -\nu r ^{\nu-2}\tan(\theta/2)\;,
\end{equation}
where $\Omega_{\rm F}$ is the nonconstant distribution of the magnetic field angular velocity \citep{blandford1977electromagnetic, mckinney2007disc, beskin2009mhd} applied to the parabolic fields,
\begin{equation}\label{eq:OmegaF}
    \frac{\Omega_{\rm F}}{\Omega_{\rm H}} = \frac{\sin{^{2} \theta_{\rm H}}[1+\ln{\mathcal{G}}]}{4\ln{2}+\sin{^{2} \theta_{\rm H}}+[\sin{^{2} \theta_{\rm H}} - 2\mathcal{G}]\ln{\mathcal{G}}},
\end{equation}
where $\Omega_{\rm H} = a/( 2 r_{\rm H} )$ is the angular velocity of the black hole, $r_{\rm H} = 1+\sqrt{1-a^2}$ is the horizon radius, $\theta_{\rm H}$ is the polar angle on the horizon, and $\mathcal{G} = (1 + \cos{\theta_{\rm H})}$ with the values varies from $\Omega_{\rm F} = 0.5\Omega_{\rm H}$ at $\theta_{\rm H} = 0$ (on the rotational axis of the event horizon) to $\Omega_{\rm F} \approx 0.265\Omega_{\rm H}$ at $\theta_{\rm H} = \pi/2$  (at the equatorial plane on the event horizon).
With Equations (\ref{eq:bphi})-(\ref{eq:bth}) The magnetic field line is constructed by
\begin{equation}\label{eq:Bline}
    \dfrac{dr}{B_{r}}=\dfrac{d\theta}{B_{\theta}/r}=\dfrac{d\phi}{B_{\phi}/(r\sin\theta)}\;.
\end{equation}

\subsubsection{Dynamical Structure}\label{sec:model_dynamics}
Although the force-free assumption neglects plasma inertia, an effective approach to estimating fluid velocity involves employing the concept of drift velocity \citep{narayan2007}:
\begin{equation}\label{eq:velocity}
    \textbf{v} 
    = R\Omega_F\dfrac{B_{p}^{2}}{B^{2}}\hat{\phi} - R\Omega_F \frac{B_{\phi}}{B^{2}}\textbf{B}_{p}\;,
\end{equation}
where $B_{p}$ is the poloidal magnetic field and $B^{2}=B_p^2+B_\phi^2$ is the square of magnetic field strength. The pathline with a velocity described by Equation (\ref{eq:velocity}) can be determined by $dx^{a}/dt=v^{a},~~a=(r, \theta, \phi)$.
In accordance with causality, the position of the light cylinder, $R_{\rm LC}\equiv c/\Omega_{\rm F}$, defines the boundary beyond which the flow becomes predominantly poloidal. In the region $R\gg R_{\rm LC} $, $B \sim |B_{\phi}|$ and $v_{p}>v_{\phi}$; in the region $R\ll R_{\rm LC} $, $B \sim |B_{p}|$ and $v_{\phi}>v_{p}$. These properties play a crucial role in governing the features of the parsec-scale jet, in regions where $R\gg R_{\rm LC}$ is satisfied.

\begin{figure}
    \centering
    \includegraphics[width =0.5\textwidth]{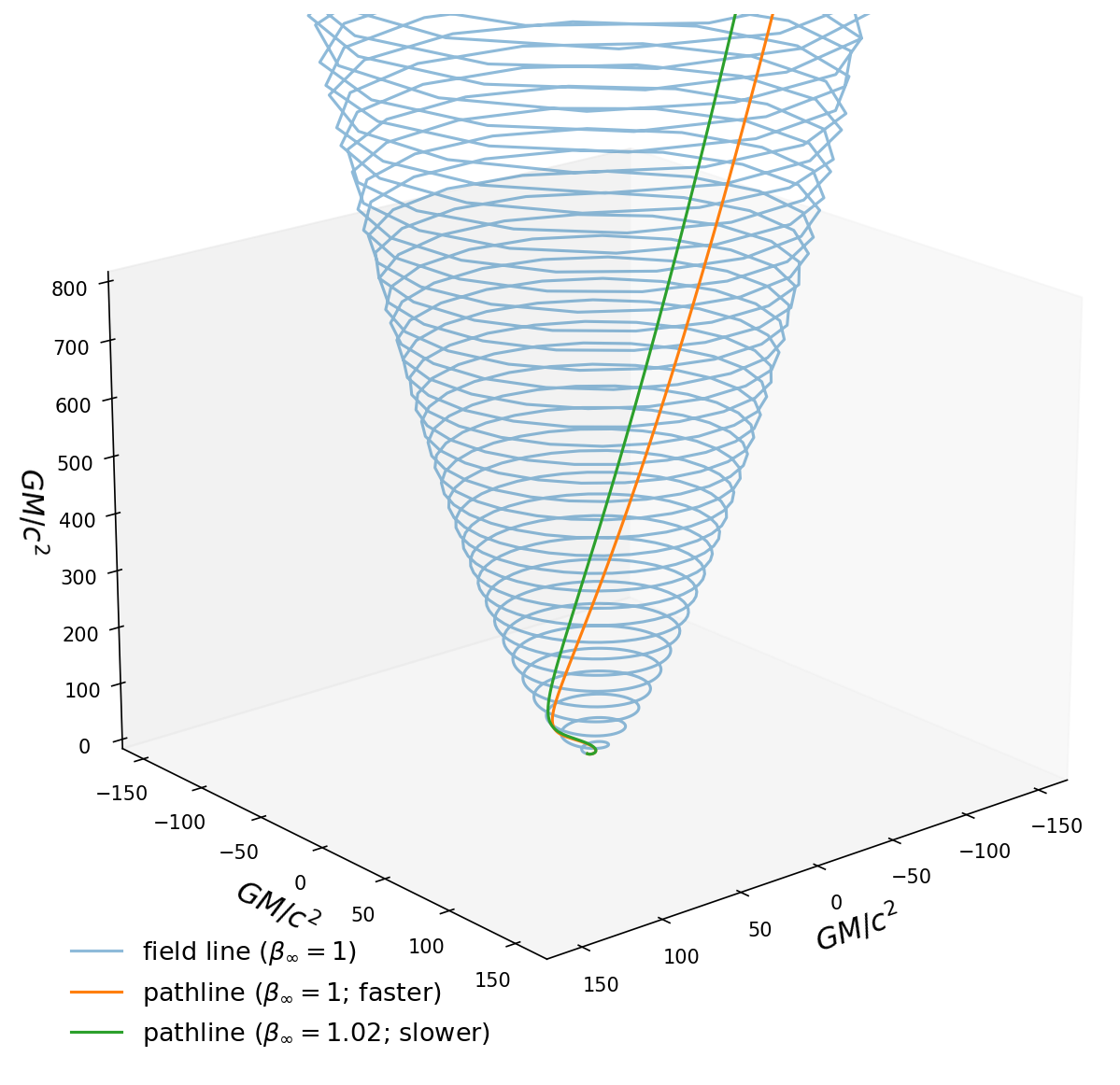}
    \caption{Example magnetic field line (blue) and pathlines (orange and green) for plasmoids flow along the field line defined by the same streamline function. The magnetic field configuration and the two pathlines are all computed from the same locations, with different values of $\beta_{\infty}$ in Equation (\ref{eq:bphi_new}), as indicated in the label. }
    \label{fig:pathline}
\end{figure}

\subsection{Symmetric mass loading and Asymmetric Plasmoid}\label{sec:mass_loading}
Observational data indicate that jet emissions exhibit stationary and time-dependent features \citep[e.g.][]{walker2018structure}. To account for these characteristics, we examine two types of mass loading within the jet model: steady, axisymmetric mass loading and intermittent, asymmetric plasmoids that occur randomly within the jet. The former establishes a stationary background for jet emissions, while the latter superimposes time-dependent emission features onto this background. For both cases, Equation (\ref{eq:velocity}) is assumed to provide a valid approximation.

In our jet model, the steady, axisymmetric mass loading
accounts for the limb-brightened jet emission. 
\citet{takahashi2018fast} proposed that such a feature can be a consequence of a relatively larger mass loading near the jet boundary, with an explicit form of the electron number density $n_{e}$   defined at a certain height $z = z_1$:
\begin{equation}\label{eq:massloading}
    n_{e}(R, z_1) = n_0\exp{\left(-\frac{(R-R_p)^2}{2\Delta^2}\right)}\;,
\end{equation}
where $R_p$ is the injection location (radius) where $n_{e}$ peaks on the plane, $\Delta$ gives the width of each peak, and $n_0$ is a normalization constant. Equation (\ref{eq:massloading}) is revised from \citet{broderick2009imaging} by substituting a single Gaussian mass loading ($R_{p}=0$) with dual Gaussian mass loadings ($R_{p}\neq0$). Above the height, $n_{e}$ along each streamline is determined by the continuity equation. 

In comparison, intermittent and asymmetric plasmoids are assumed to be injected within the jet at arbitrary locations (specified by the pathline parameters, $\theta_{\rm H}$ and $\phi_{\rm H}$, and the location on the pathline), with its radius  $dr_{\rm inj}$,
and the emissivity higher than the background value with $dj_{\rm inj}$ times. 
An additional parameter $dl_{\rm inj}$ is also introduced to mimic an elongated shape of the plasmoid, possibly resulting from the shearing effect (see \S \ref{sec:asym_inj}).

\begin{figure*}[ht]
    \centering
    \includegraphics[width =0.9\textwidth]{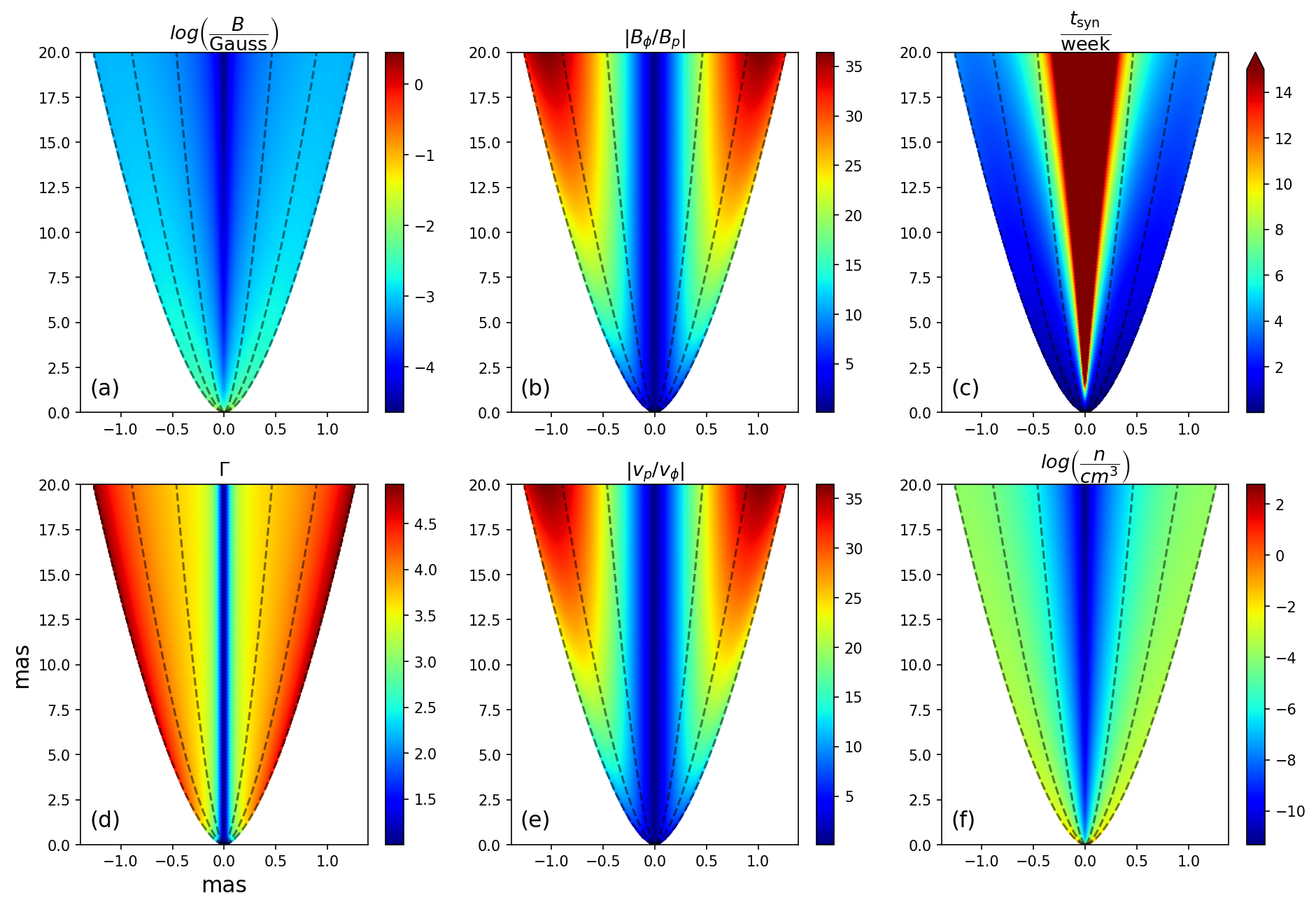}
    \caption{Model jet structure and properties. (a) The color map of the magnetic field $B$. (b) The ratio of toroidal and poloidal components of the magnetic field strengths. (c) The cooling time scale. (d) Lorentz factor. (e) The ratio of poloidal and toroidal components of the velocity field strengths. (f) The number density distribution of the nonthermal electrons. The dashed lines in each plots indicate the streamline function $\Psi$ which attaches the black hole horizon at $\theta_{\rm H}=(30^{\circ},60^{\circ},90^{\circ})$. The light cylinder $R_{\rm LC}$ is located around $0.02\sim0.04$ mas.}
    \label{fig:model_overview}
\end{figure*}

\subsection{Model Setup for M87 Jet}\label{sec:model_setup}
Taking into account the aforementioned force-free jet structure and mass loading, we subsequently construct our jet model for M87, incorporating the subsequent considerations.

The global magnetic field configurations in radiatively inefficient accretion flows, such as in the case of M87, have been investigated in several numerical simulations \citep[e.g.][]{hirose2004magnetically,mckinney2004measurement}. These studies have demonstrated that the jet can be powered by the magnetohydrodynamical extraction of the black hole's rotational energy through a large-scale magnetic field threading the horizon. Consequently, we consider the global magnetic field of the jets as large-scale magnetic field lines that are connected to the black hole event horizon, and adopt the streamline attached to the event horizon at the equatorial plane as the outer boundary of the jet \citep[c.f.][]{takahashi2018fast}. This conceptual framework is supported by the observations of M87 that the jet width of M87 can be fairly fitted by Equation (\ref{eq:streamline}), with $\nu=0.75$ \citep{asada2012,nakamura2018}.

It is well known that the jet velocity predicted by the force-free jet model is actually faster than the M87 jet velocity inferred by observations \citep[e.g.][]{nakamura2018}.  An alteration in the jet dynamics to correspond with a slower terminal  Lorentz factor can be achieved by diminishing the toroidal magnetic field, As introduced in \citet{lu2014}, with  
\begin{equation}\label{eq:bphi_new}
    \textbf{B}^{\rm modified}_\phi = \beta_\infty(\psi) \textbf{B}_\phi\;,
\end{equation}
where the factor $\beta_{\infty}\geq1$ is related to the terminal Lorentz factor $\Gamma_{\infty}$ with $\beta_{\infty}=(1-\Gamma_{\infty}^{-2})^{1/2}$. This adjustment is theoretically justified by the influence of the plasma on the force-free magnetic field, thereby affecting the magnetic field pitch angle and jet velocity \citep[e.g.][]{pu2020,pu2022modeling}. 
Note that the toroidal velocity also decreases with a reduction in the poloidal terminal velocity, and therefore with a higher value of $\beta_{\infty}$.
To illustrate the aforementioned characteristics, Figure \ref{fig:pathline} provides example pathlines for the velocity defined by Equation (\ref{eq:velocity}), with different choices of $\beta_{\infty}$. For comparison, example 3D magnetic field lines, computed using Equations (\ref{eq:bphi})-(\ref{eq:bth}) and (\ref{eq:Bline}), are also presented.  It is shown that, on the subparsec scale of the M87 jet ($\approx 10^{3} R_{\rm g}$), the pathline is mostly poloidal as it is already far away from the light cylinder.

 \begin{figure*}
  \centering
\subfigure{\includegraphics[width=0.95\textwidth]{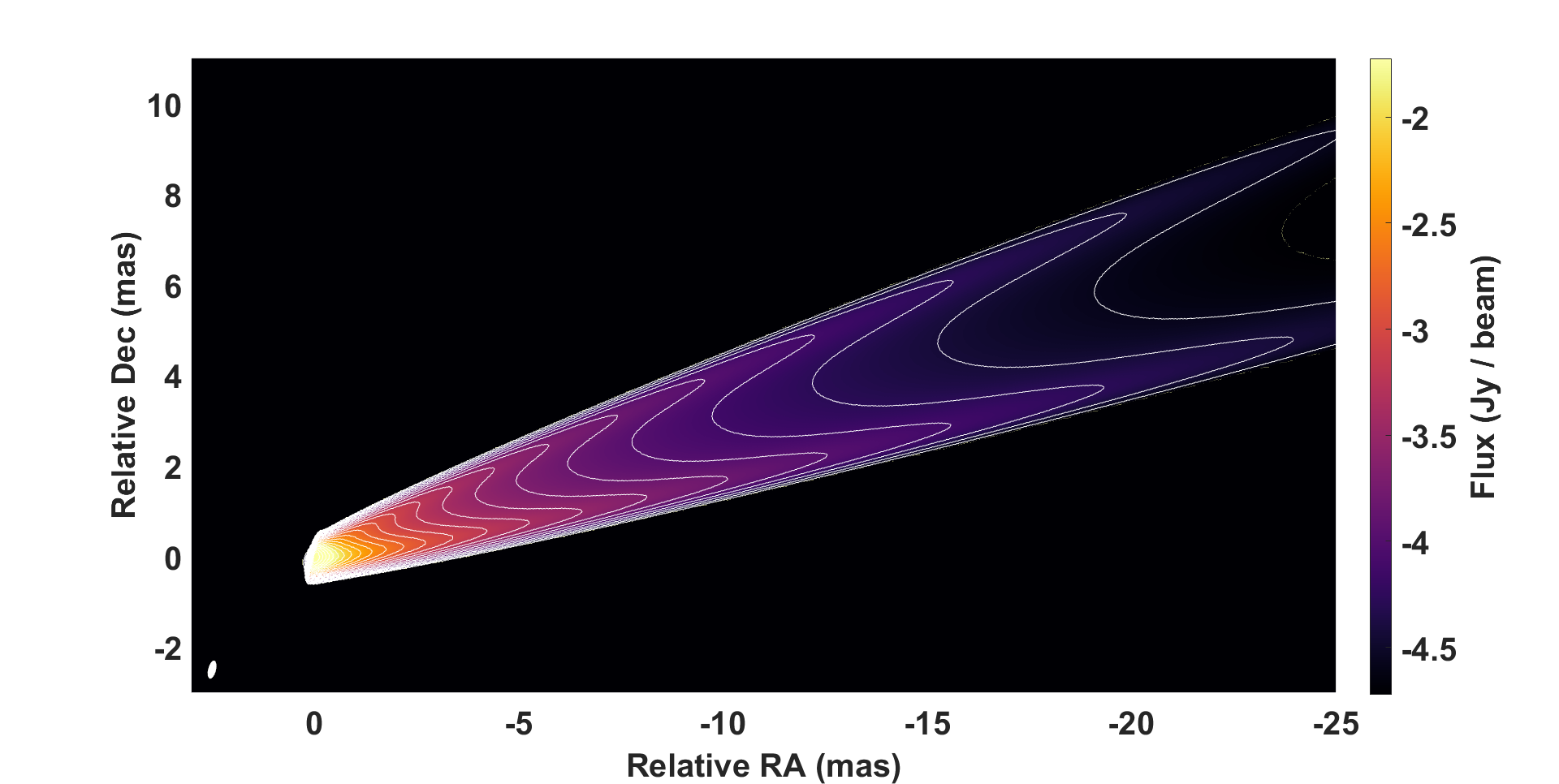}}

  \caption{
  Blurred model M87 jet image at 43 GHz.  The contour levels start at a maximal value per beam, then decrease in steps of factors of 2. While the forward jet is obvious, the counterjet is also much fainter due to relativistic beaming. The image is convolved with a gaussian beam size of 0.43 mas $\times$ by 0.21 mas, elongated along the position angle $-16^{\circ}$, to mimic the observation \citep{walker2018structure}.}
  \label{fig:jet_image_bg}
\end{figure*}

Whereas a constant $\beta_{\infty}(\psi)$ is assumed in \citet{lu2014}, this study considers a variable $\beta_{\infty}(\psi)$ in terms of $\theta_{\rm H}(\psi)$, of the form \begin{equation}\label{eq:beta_vary}
    \beta_{\infty}(\theta_{\rm H}(\psi))=A+(B-A)\cos(\theta_{\rm H}) \;,
\end{equation} where the coefficients $A=1.02$, $B=1.04$, and $0\leq\theta_{\rm H}(\psi)\leq \pi/2$ represent the latitude at which the streamline $\psi$ attaches to the event horizon, $r=r_{\rm H}$. The coefficient $A$ is determined by adopting $\Gamma_{\infty}\approx 5$, and $B$ is chosen to be a value slightly larger than $A$ to mimic the slower spine and fast sheath jet dynamics \citep[e.g.][]{mckinney2004measurement}. 

We adopt the distance $D=16.8$ Mpc and black hole mass $M_{\rm BH}=6.5\times 10^{9}M_{\odot}$. The dimensionless black hole spin parameter $a=0.95$, and the orientation of the black hole's spin axis are directed away from Earth \citep{2019ehtV}, forming an angle of $i\approx200^{\circ}$ with the observer's line of sight. 

\citet[][]{takahashi2018fast} found that, in contrast to a slowly rotating black hole, a rapidly spinning black hole demonstrates a more symmetrical limb-brightened jet. This phenomenon arises because the light cylinder is positioned closer to the black hole, thus diminishing the influence of the asymmetric limb-brightened jet feature caused by $v_{\phi}$ (see also \S \ref{sec:model_dynamics}). In \citet[][]{takahashi2018fast}, the outer boundary of the jet extends beyond the streamline associated with the event horizon on the equatorial plane. We notice that the above reason still holds for our model, where the outer boundary of the jet is defined by the last large-scale magnetic field lines, which attach onto the event horizon at the equatorial plane, resulting in a similar dependence of spin on the symmetry of the limb-brightened jet. Consequently, we select a high black hole spin case $a=0.95$ as our fiducial parameter.

In the next section, we estimate the apparent motion of the plasmoid and compute the resulting image of the asymmetric plasmoid onto the background stationary jet. 
The jet model images are computed by integrating the nonthermal synchrotron emission, with the emissivity and absorptivity given in \citet[][]{pandya2016polarized}. The energy distribution of nonthermal electrons is described by a power-law distribution with index $p=2.0$, and cutoffs $\gamma_{\rm min}=100$ and $\gamma_{\rm max}=10^{6}$.
We modify the GPU-based radiative transfer code \citep{pu2016odyssey} for these radiative transfer computations in flat spacetime. The distribution of nonthermal electrons, Equation (\ref{eq:massloading}), is specified by choosing $z_{1}=5~R_{\rm g}$ and $R_{p}=4.435~R_{\rm g}$, with  $n_{0}$  determined by a total flux $\sim 0.65$ Jy with a field of view with $\sim 0.15'' $ radius at 43 GHz. The normalization of magnetic field strength is considered by assuming that the magnetic field strength is $B\sim 10^{2}$ G near the core and decays with the distance \citep{ro2023spectral}.

Figure \ref{fig:model_overview} displays some properties of our jet model computed with the above mentioned normalization of magnetic field strength and number density, including magnetic field strength $B$, magnetic pitch angle $|B_{\phi}/B_{p}|$, ratio between poloidal and toroidal velocity\footnote{From Equation (\ref{eq:velocity}), $|v_p/v_{\phi}|=|B_{\phi}/B_{p}|$.} $|v_p/v_{\phi}|$, Lorentz factor, and mass loading. In addition, the estimated synchrotron cooling time scale \citep{tsvi04},
\begin{equation}\label{eq:tcool}
    t_{\rm syn}=\dfrac{3}{\sigma_{\rm T}}\sqrt{\dfrac{2\pi c m_{e} q_{e}}{B^{3}\Gamma}}\nu^{-1/2}\;,
\end{equation}
is also shown in panel (c), by applying $\nu$ = 43 GHz in the above Equation. 
In Equation (\ref{eq:tcool}), $\sigma_{\rm T}$ is the Thompson cross section, $m_{e}$ and $q_e$ are the mass and charge of the electron, respectively. The energy distribution of relativistic electrons does not appear explicitly in the formula, as it is expressed through the critical frequency of synchrotron radiation and absorbed in the frequency dependence \citep[see][]{tsvi04}.  Due to a smaller magnetic field strength and Lorentz factor, the synchrotron cooling time scale is longer near the central region of the jet. 

\section{Comparison with observed 43 GHz Jet properties}\label{sec:comparion}
\subsection{43 GHz Background Jet morphology}

For our interest, to compare with observations at the same frequency \citep{mer2016, walker2018structure}, we are especially interested in the 43 GHz jet properties and the comparison between our model and the observations.
The 43 GHz jet image with axisymmetric mass loading, Equation (\ref{eq:massloading}), is presented in Figure \ref{fig:jet_image_bg}. Exhibiting limb-brightened jet features, the model jet morphology is visually similar to observation \citep[][see their Figure 1]{walker2018structure}. 

With the assumed viewing angle $i=200^{\circ}$ between the black hole's rotational axis and the observer's line of sight, the black hole's rotational axis is pointing away from Earth, resulting in the southern limb being brighter than the northern limb. In the following, we refer to Figure \ref{fig:jet_image_bg} as the background jet image and consider injections of asymmetric plasmoids within the jet in the subparsec jet region.

\begin{figure}[ht]
    \centering
    \includegraphics[width =0.45\textwidth]{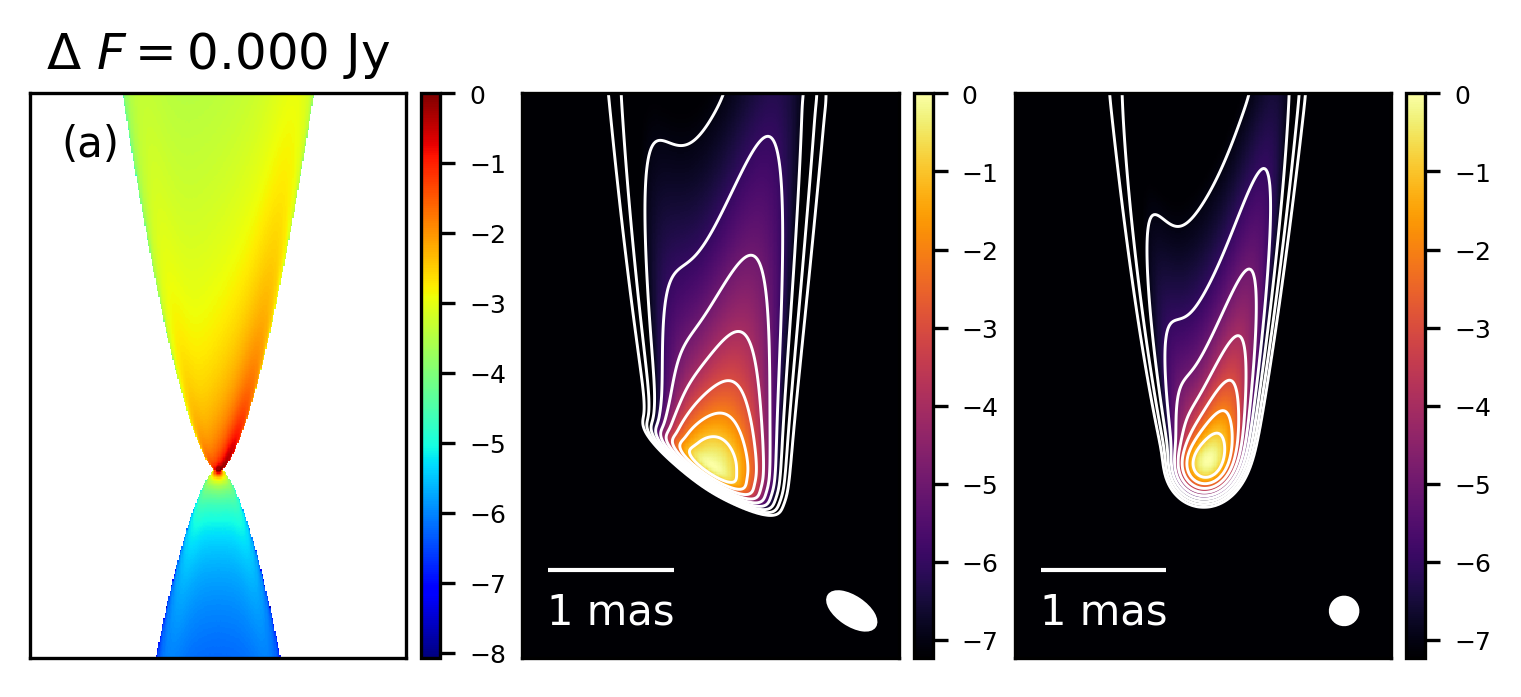}\\
      \includegraphics[width =0.45\textwidth]{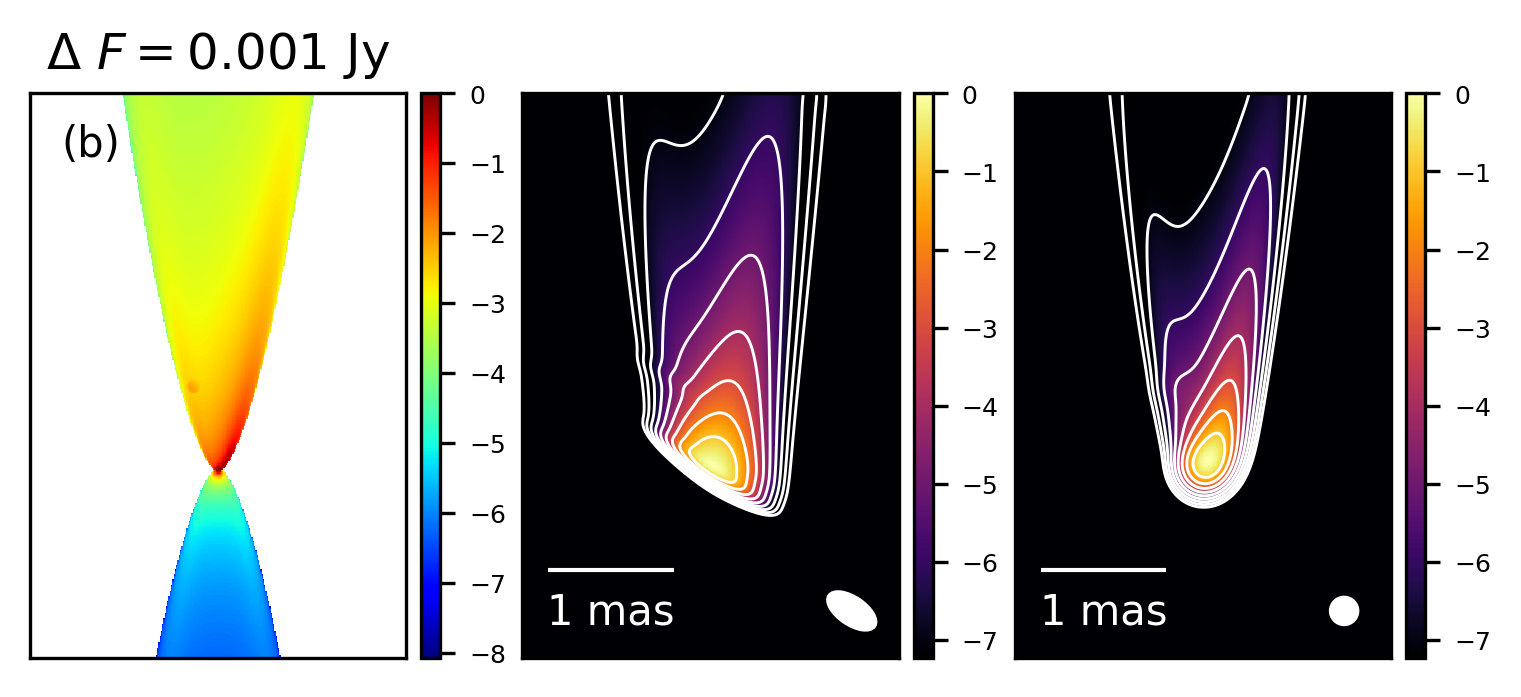}\\
    \includegraphics[width =0.45\textwidth]{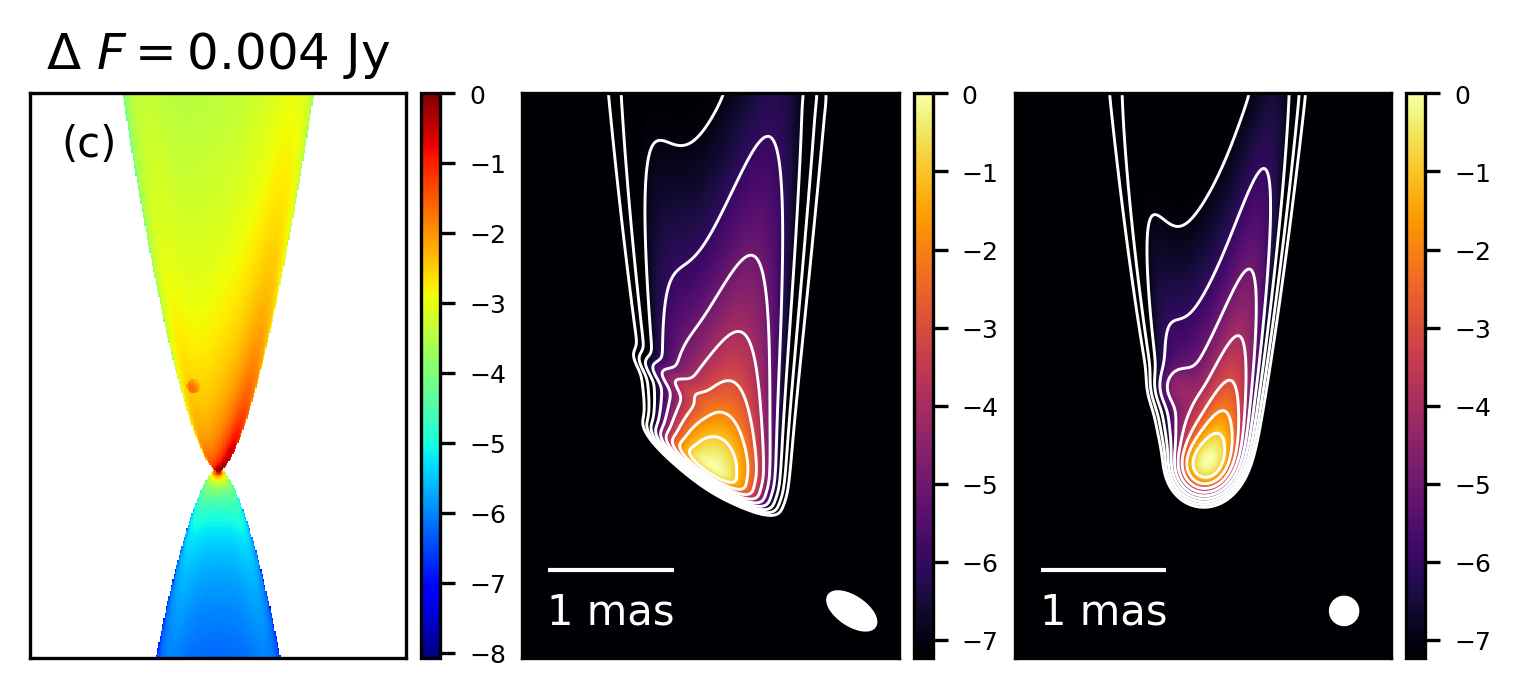}\\
    \includegraphics[width =0.45\textwidth]{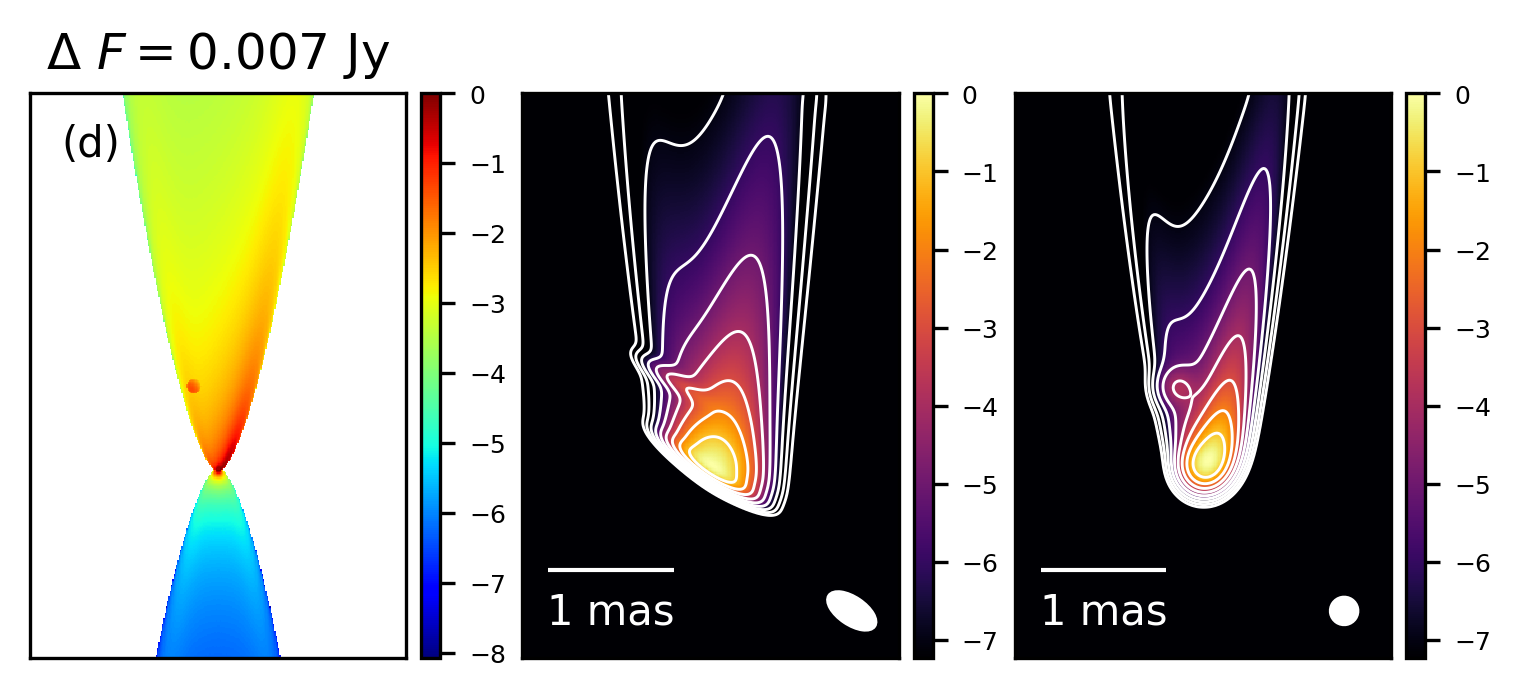}\\
    \caption{Model images with a spherical plasmoid of
    different brightness $dj_{\rm inj}$.  The radius of the plasmoid is $dr_{\rm inj}=15R_{\rm g}$, with the plasmoid's emissivity assumed to be: (b)$dj_{\rm inj} =20$, (c)$dj_{\rm inj} =80$, and (d)$dj_{\rm inj} =160$. For each case, the left panel displays the computed image (in logarithmic scale), and the right panel shows a blurred image with elliptical or circular beams, with a dynamical range equal to 150.  The colorbar indicates the flux density normalized to the maximum value in each individual panel. The size of the elliptical beam is the same as in Figure \ref{fig:jet_image_bg}. 
     The beam sizes are denoted at the bottom right corner of each image. The flux difference $\Delta F$ is computed compared to the background image with $\Delta F=0$, as shown in (a). }
\label{fig:spot_sphere}
\end{figure}

\begin{figure*}[ht]
    \centering
\includegraphics[width =0.45\textwidth]{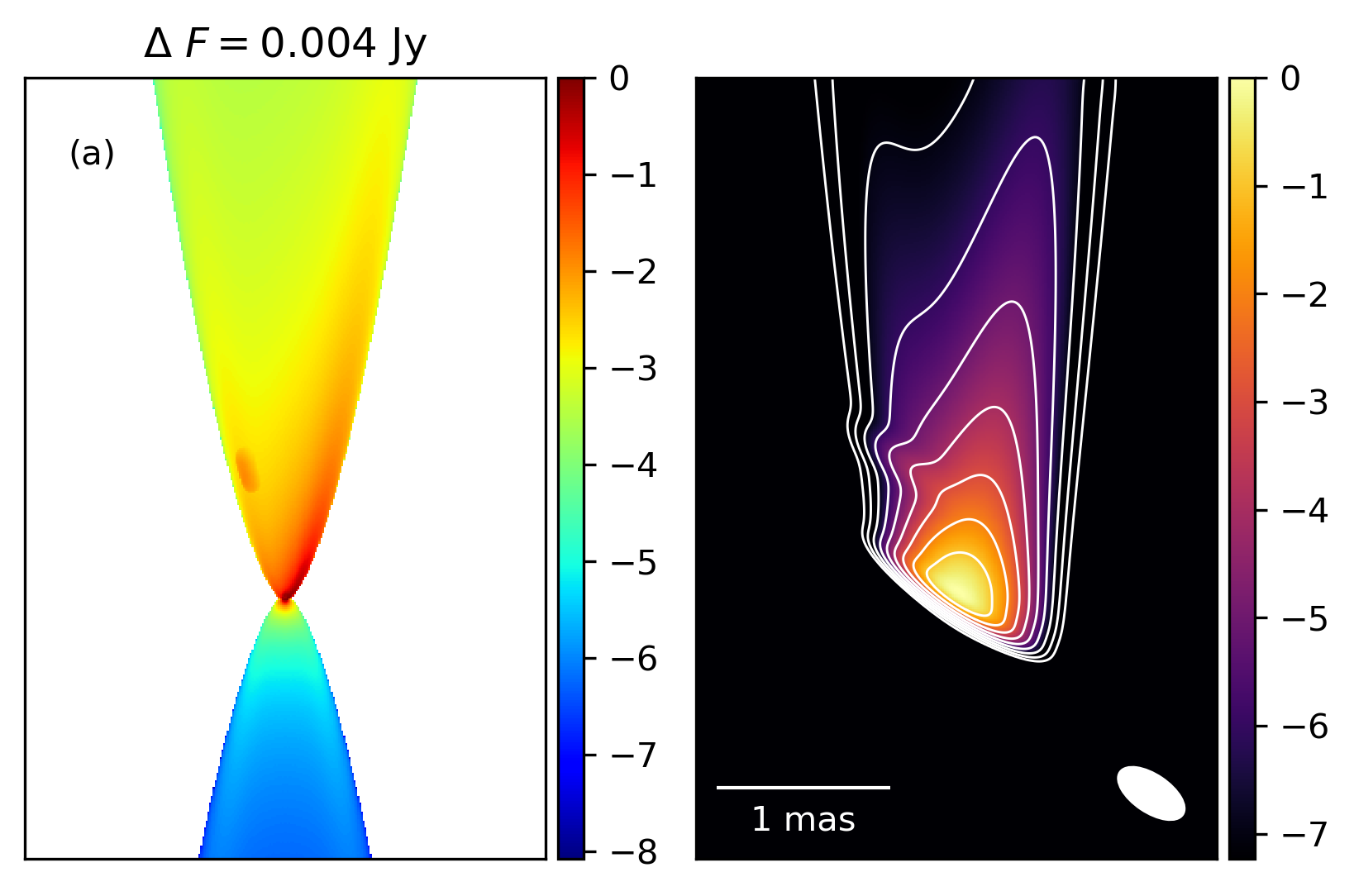}
            \includegraphics[width =0.45\textwidth]{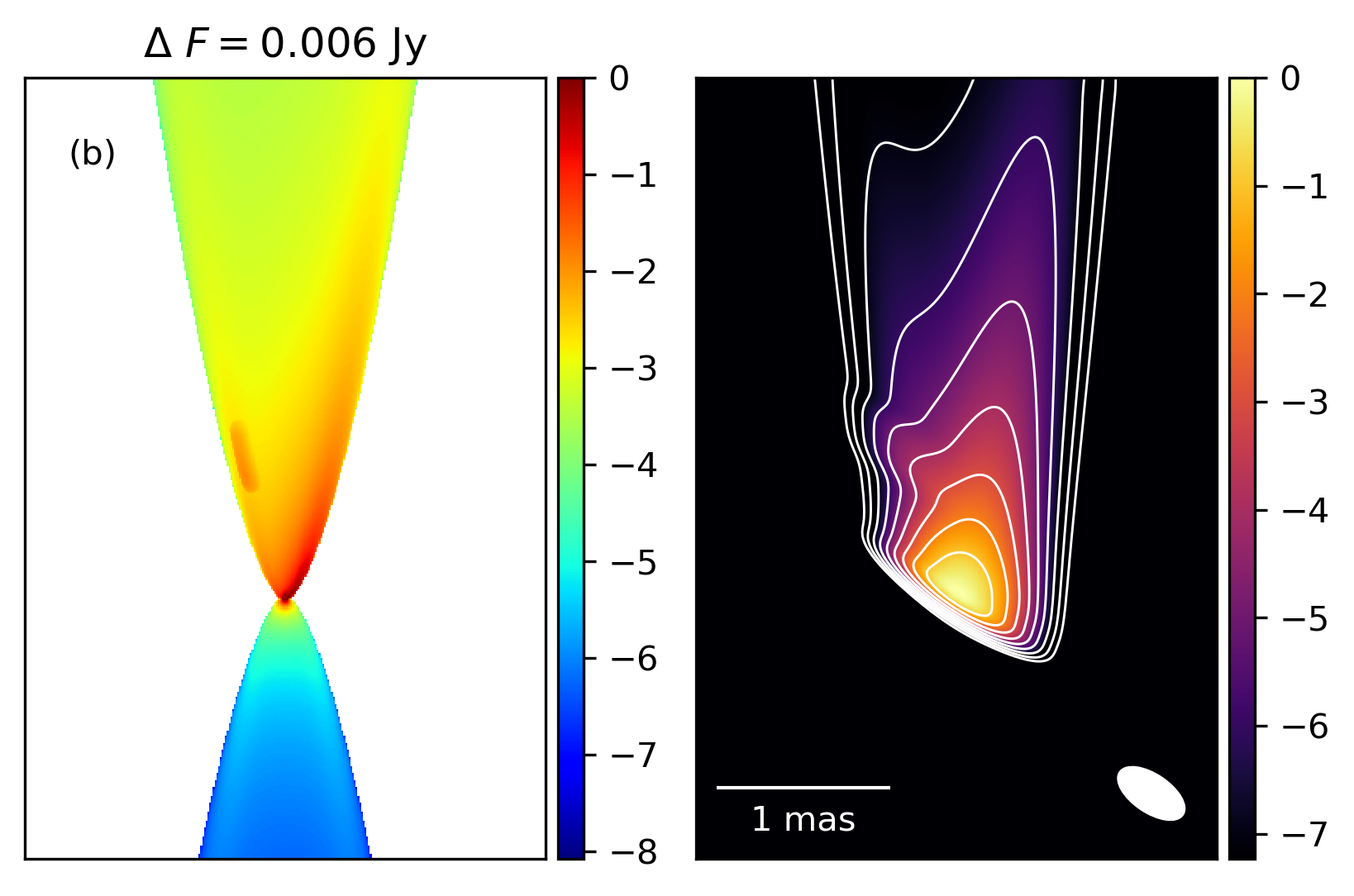}\\
    \includegraphics[width =0.45\textwidth]{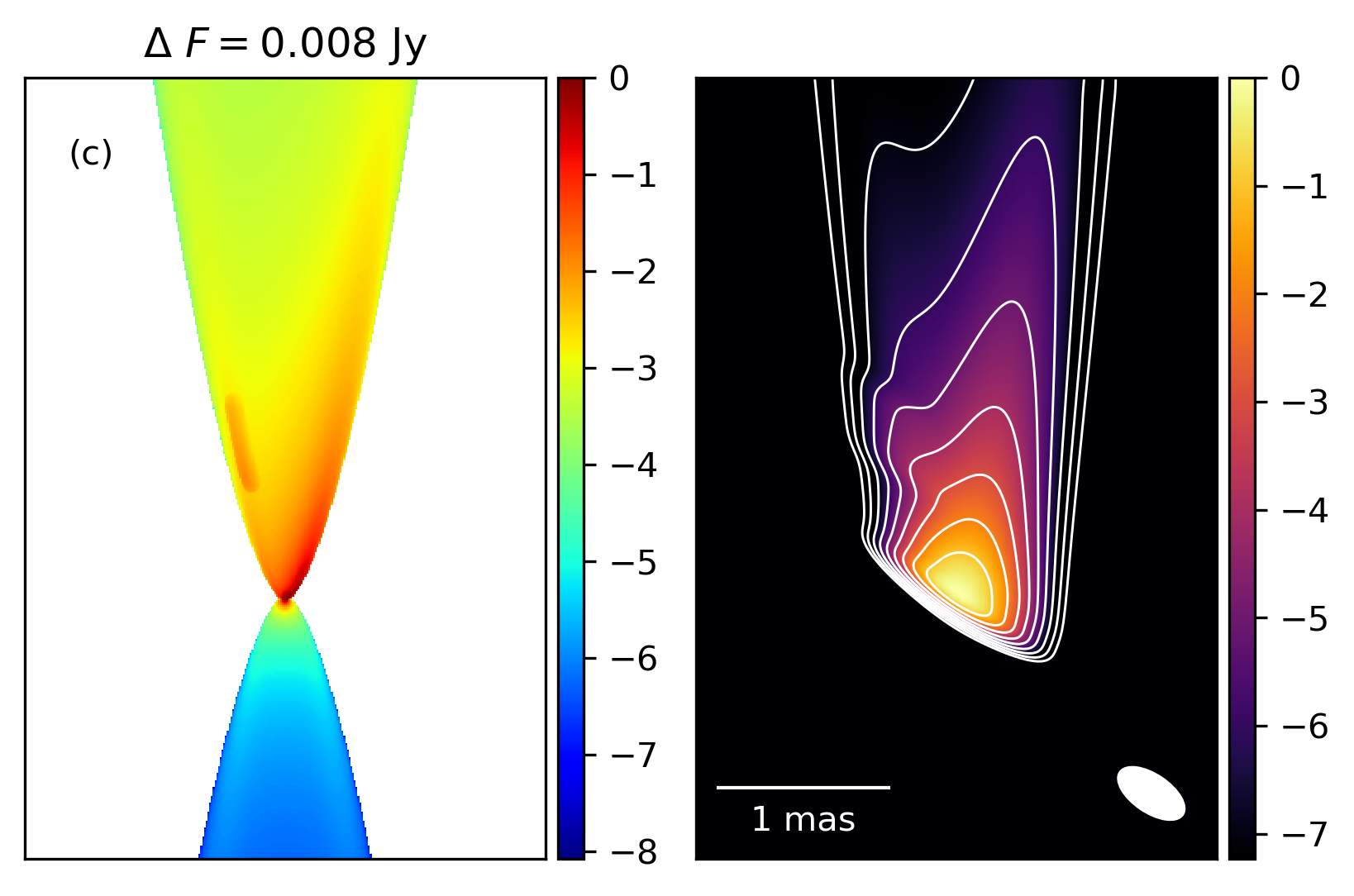}
    \includegraphics[width =0.45\textwidth]{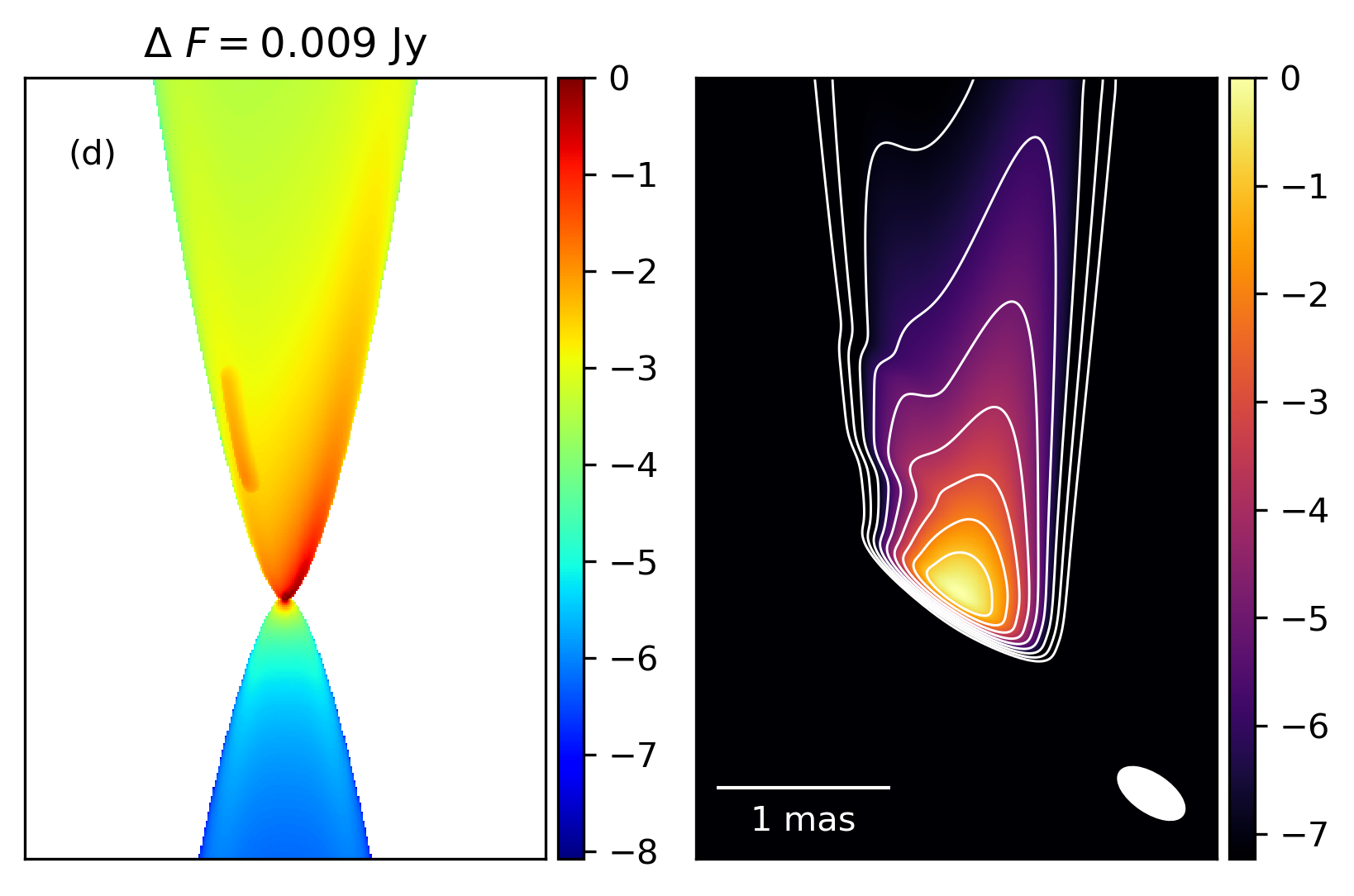}\\        
    \includegraphics[width =0.45\textwidth]{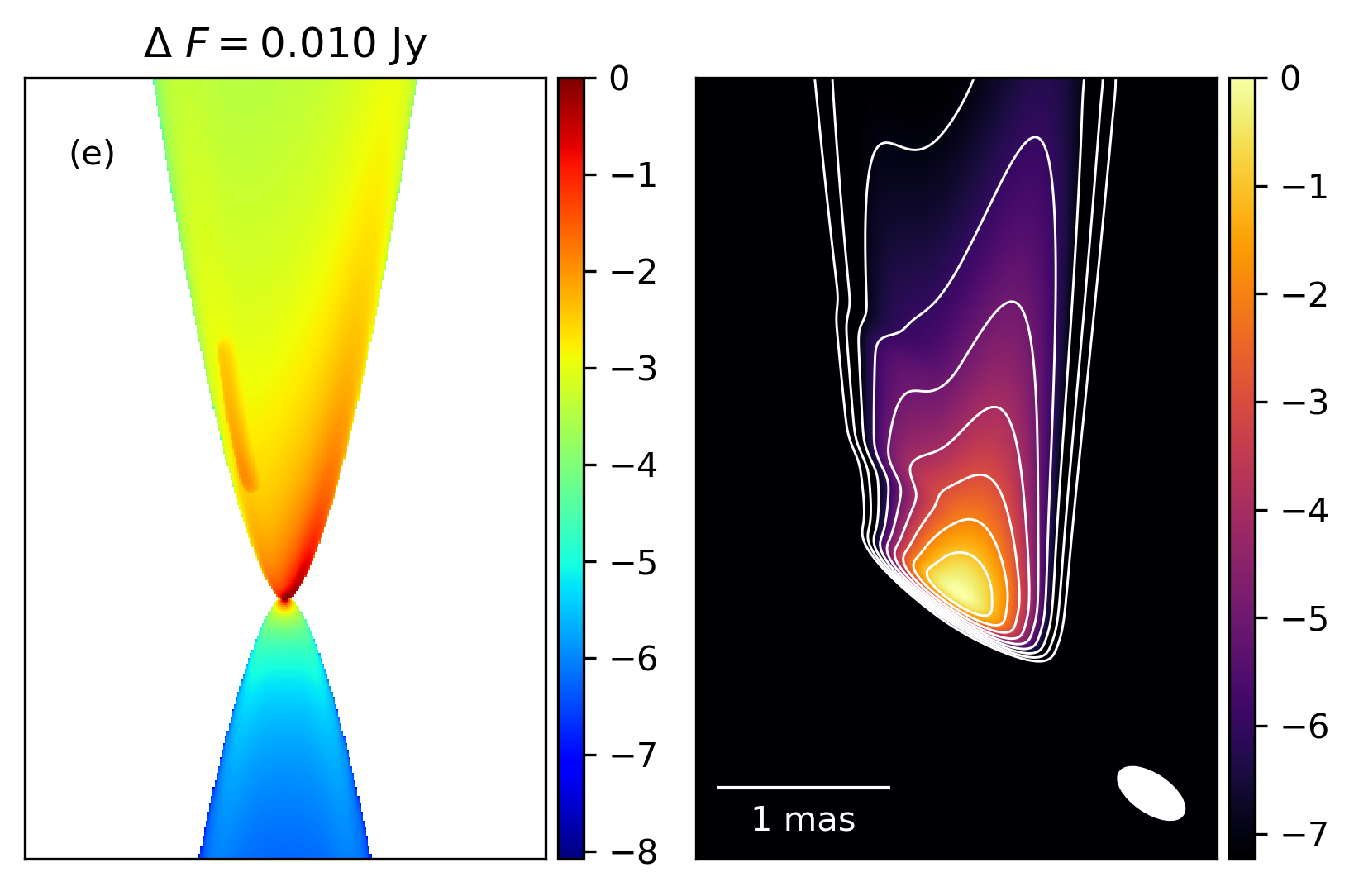}
    \includegraphics[width =0.45\textwidth]{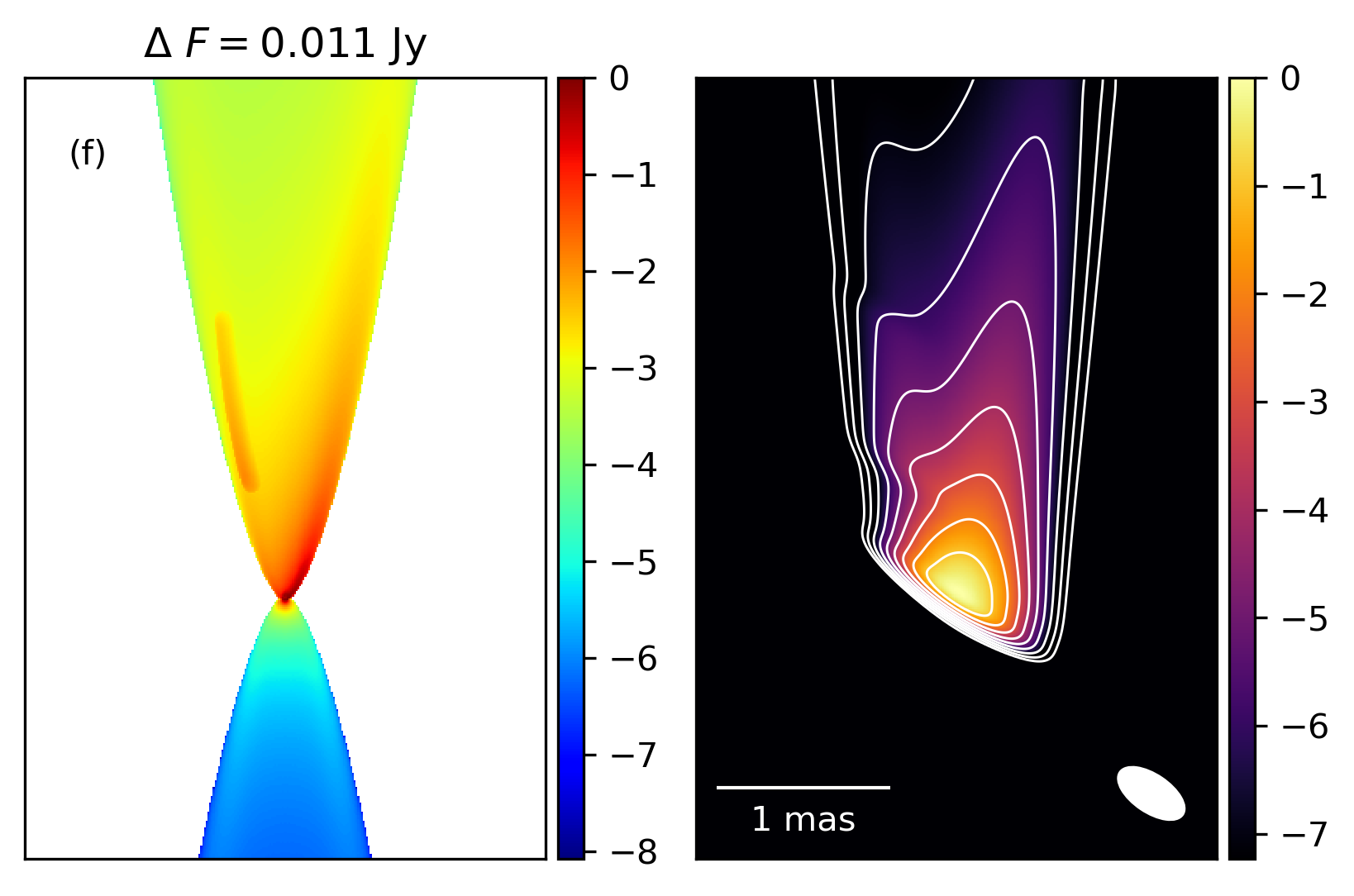}
    \caption{Model images with asymmetric shearing plasmoid injected within the background jet. 
    The lengths of the shearing plasmoid are characterized by varying length $dl_{\rm inj}=(100,200,300,400,500,600)R_{\rm g}$, from (a) to (f). 
    Other parameters are assumed to be 
    $dr_{\rm inj}=15R_{\rm g}$ and $dj_{\rm inj}=20$. The flux difference $\Delta F$ is computed compared to the background image shown in Figure \ref{fig:spot_sphere}(a). }    \label{fig:asym_spot_left}
\end{figure*}

\begin{figure*}[ht]
    \centering
      \includegraphics[width =0.45\textwidth]{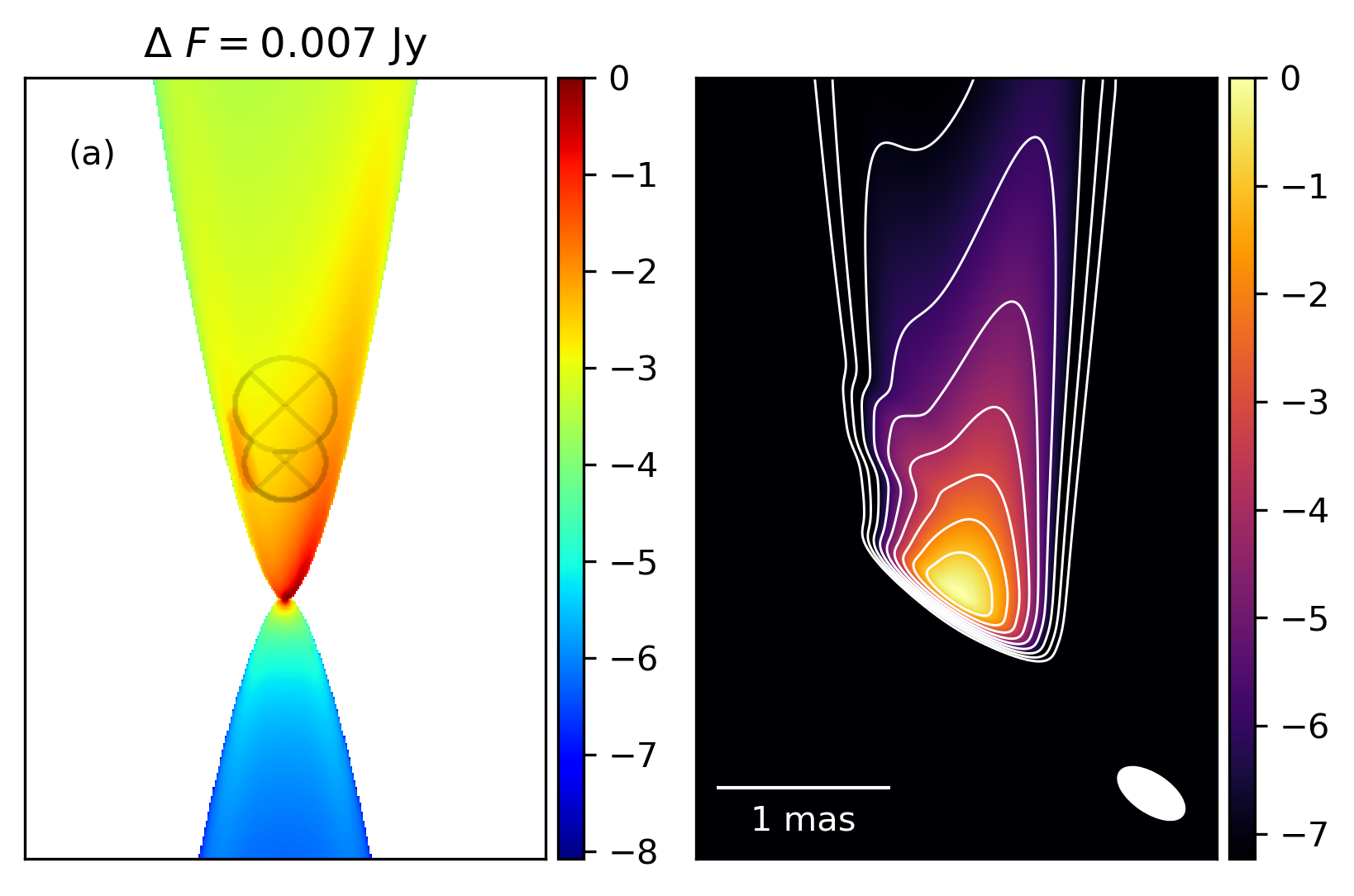}
    \includegraphics[width =0.45\textwidth]{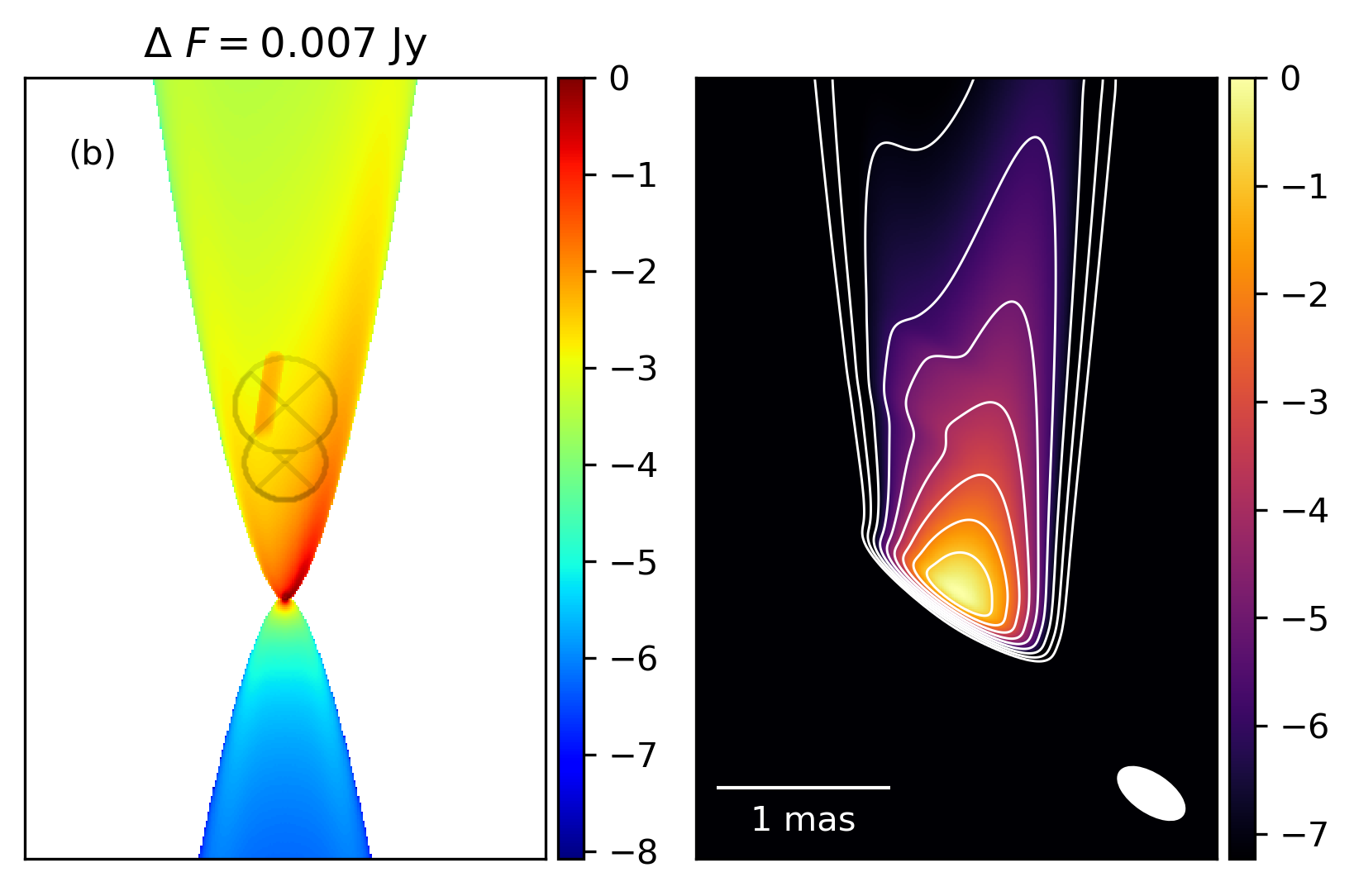}\\
      \includegraphics[width =0.45\textwidth]{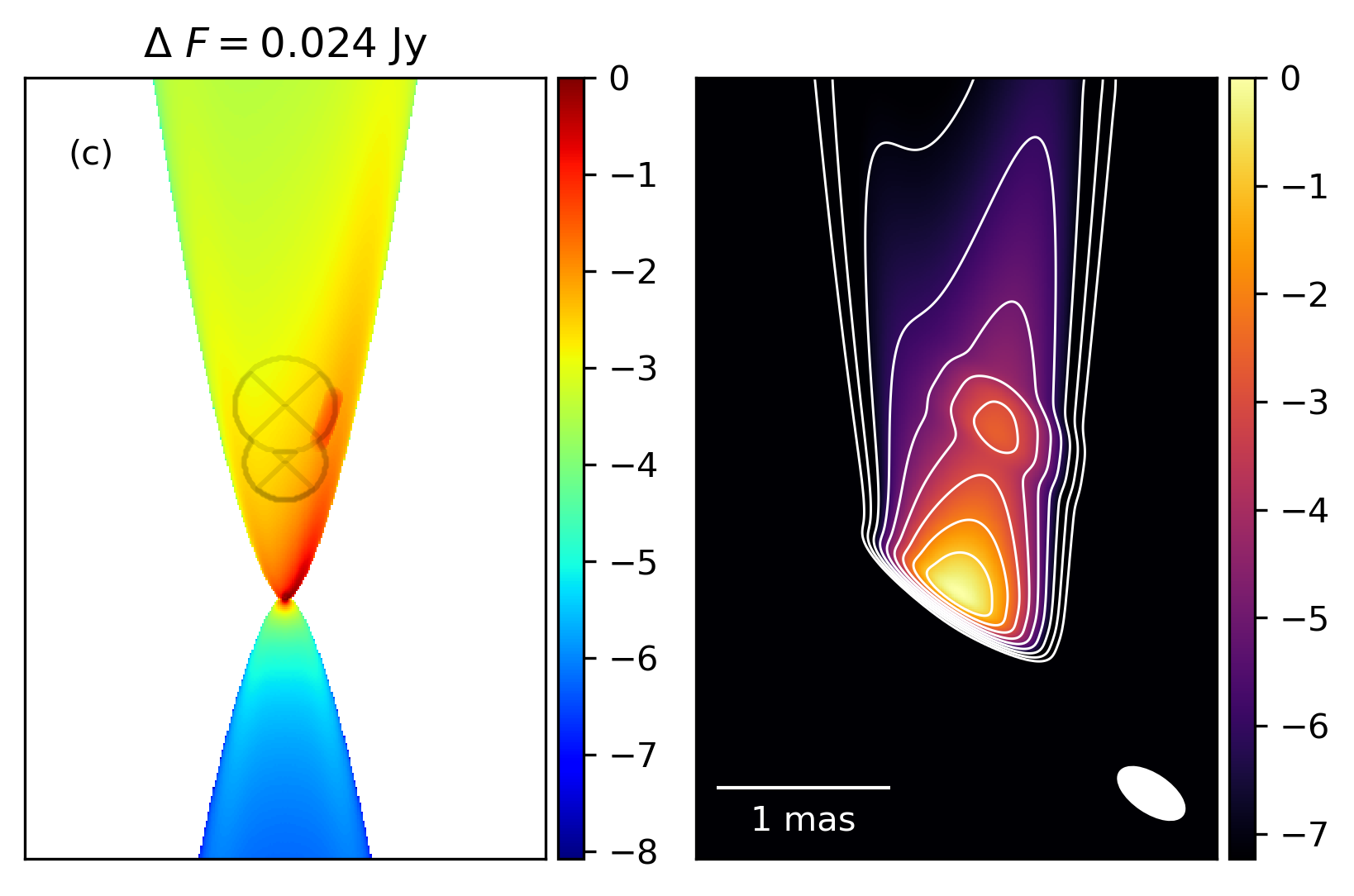}
    \includegraphics[width =0.45\textwidth]{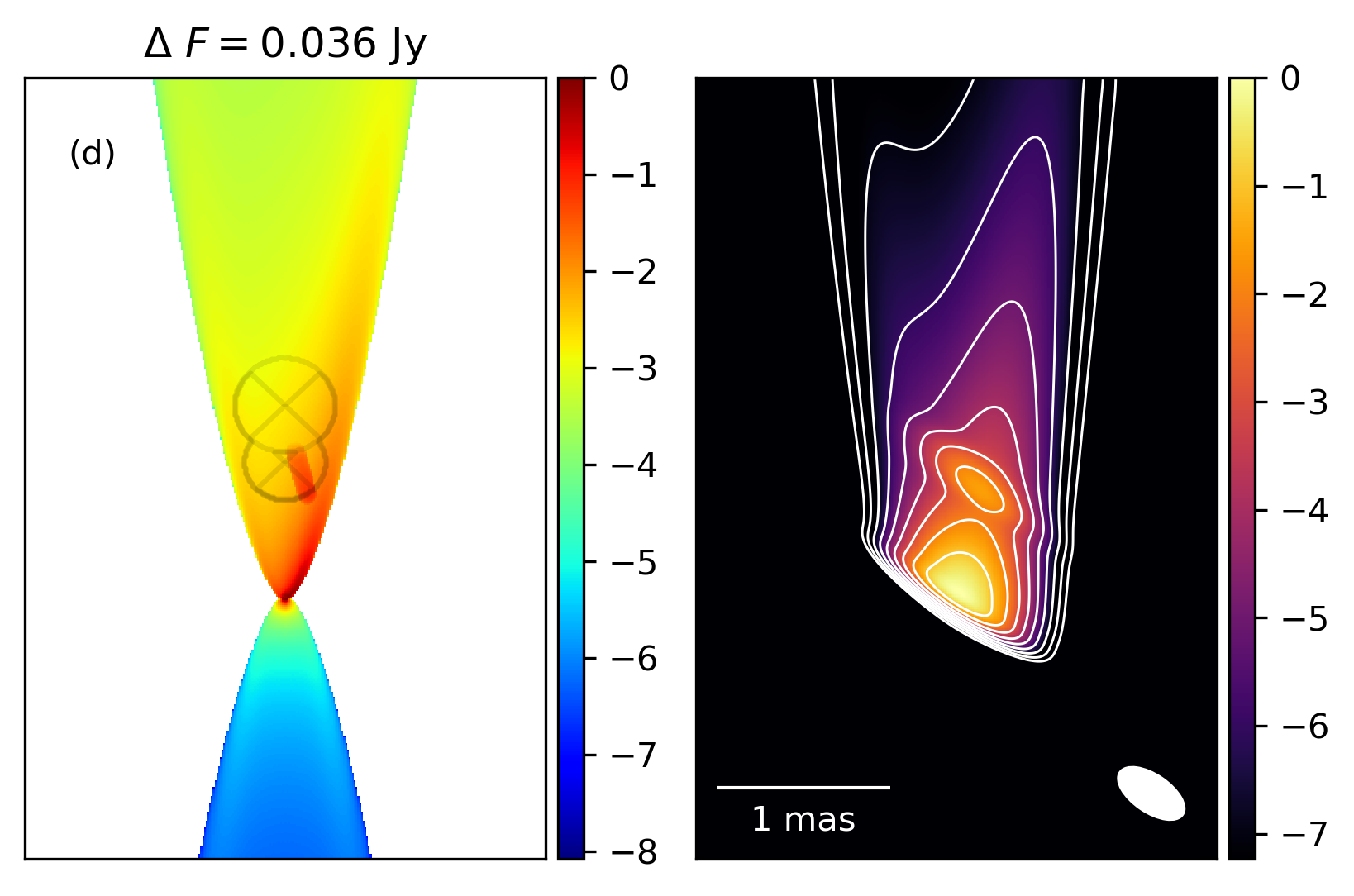}\\
    \caption{Model images shearing plasmoid injected at different azimuthal angles. The shearing plasmoids are located at different azimuthal angles along the pathlines defined by $\theta_{\rm H}=60^{\circ}$, with relative azimuthal angle
     $\Delta\phi_{\rm H} = (0^{\circ}, ~90^{\circ}, ~180^{\circ},~270^{\circ})$ for (a) to (d). 
    The parameters $dl_{\rm inj}=250~R_{\rm g}$ and $dr_{\rm inj}$ are fixed. 
     Two circular planes of constant height are overlapped onto the model images. See text for details explaining the variation of the projected length. The flux difference $\Delta F$ is computed compared to the background image shown in Figure \ref{fig:spot_sphere}(a).}
\label{fig:spot_all_around}
\end{figure*}

\subsection{Asymmetric plasmoid injection, shearing, evolution, and cooling}\label{sec:asym_inj}
The injected plasmoid within the jet can modify the jet images by adding additional features to the background image shown in Figure \ref{fig:jet_image_bg}. In general, plasmoid injection can be intermittent and asymmetric, and the plasmoid can also experience a shearing effect before it loses its energy due to synchrotron cooling. 
In order to examine the influence of injected plasmoids, which are subject to shearing, evolutionary, and cooling phenomena, on the morphology of jets, we adopt the following considerations. Noting that shearing is predominantly gorverned by poloidal velocity 
(cf. the cases when the injected plasmoids are relatively close to the black hole \citep[e.g.][]{jeter2020}), as discussed in \S \ref{sec:model_dynamics}, we approximate the size of a shearing plasmoid as the region enclosed by overlapping spheres with radii $dr_{\rm inj}$, centered along a pathline with distance $dl_{\rm inj}$. The distance $dl_{\rm inj}$ along the pathline can be interpreted as a result of the time evolution of the plasmoid after injection, with constraints from the cooling time scale. We also define $dl_{\rm inj}=0$ as the case when the injected plasmoid has a spherical shape with radius $dr_{\rm inj}$.
In addition, the emissivity of the plasmoid is parameterized by assigning the ratio $dj_{\rm inj}$ between the emissivity within the plasmoid region and the background in Figure \ref{fig:spot_sphere}.
The pathlines are computed from selected starting points, which attach the black hole horizon at different latitudes $\theta_{\rm H}$ and azimuthal angle $\phi_{\rm H}$. Without loss of generality, here we consider the pathlines characterized by $\theta_{\rm H}=60^{\circ}$.

We first examine a spherical plasmoid with varying $dj_{\rm inj}$ in Figure \ref{fig:spot_sphere}. For each case, we present the computed jet image, the blur images with an elliptical beam size similar to the observation in \citet{walker2018structure}, and with a circular beam, with the same field of view. 
The discrepancies in flux $\Delta F$ compared to the background image, particularly panel (a) with $\Delta F=0$, are specified in the title for each scenario. Figure \ref{fig:spot_sphere} demonstrates the potential variation in the observation image influenced by both the beam size and the brightness of the plasmoid. It is evident that the blurred image may lack localization of the injected plasmoid, despite the clear identification of the spherical plasmoids in the computed images.

To investigate the influence of shearing, Figure \ref{fig:asym_spot_left} illustrates a series of model images with various selections of $dl_{\rm inj}$, with fixed $dr_{\rm}=15$ and $dj_{\rm inj}=20$. In particular, we assume that the shearing spot is positioned on the dimmer side of the background image, which is achievable by selecting an appropriate $\phi_{\rm H}$.  Although there is a slight variation in the total observed flux ($\Delta F < 0.02$ Jy), an increase in $dl_{\rm inj}$ of the shearing spot positioned on the dimmer side of the background image can substantially enhance the brightening characteristic of the limb on that side, subsequently altering the observed jet morphology. 
 The same characteristic is expected in other observational frequencies, as presented in Appendix~\ref{app:multifreq_img}.
We also note that the elongated direction of the shearing spot may not be clearly identified from the blurred images, influenced by both the observation beam size and the nonuniform background images.

The shearing plasmoid, with a random injection position, is expected to be situated at various possible azimuthal locations within the background jet. To examine the influence of the injection location of the plasmoid, Figure \ref{fig:spot_all_around} illustrates the scenarios where the shearing plasmoids, with fixed parameters $dl_{\rm inj}=200~R_{\rm g}$ and $dr_{\rm} =15~R_{\rm g}$, are placed at different azimuthal locations within the jet. This is achieved by altering the pathline parameter $\phi_{\rm H}$, and we specify that the differences in azimuthal angle are $\Delta \phi_{\rm H}$ with every $90^{\circ}$.
Although the length of the shearing plasmoid, $dl_{\rm inj}$, remains constant, its projected length across each panel of Figure \ref{fig:spot_all_around} varies depending on its azimuthal position. This effect is more effectively visualized by employing two circular planes of constant height, which are approximately located at the base and the apex of the shearing plasmoid and overlapped onto the model image for each case. The stochastic nature of the injection of the plasmoid leads to a variety of outcomes in the resulting jet images, which are accompanied by only a minor variation in the total observed flux of $\Delta F < 0.05$ Jy.

\begin{figure}
\begin{center}

\includegraphics[width =0.45\textwidth]{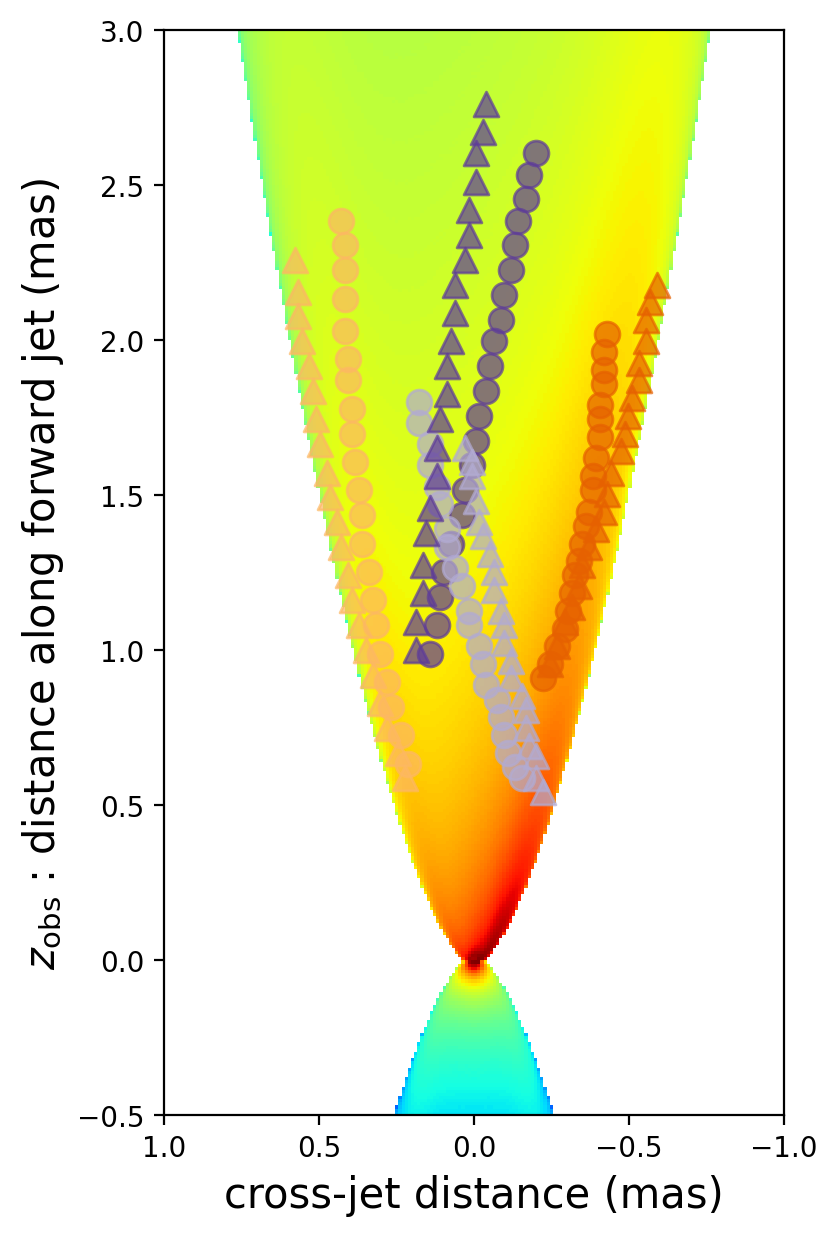}
    \caption{The trajectories of injected plasmoids, with interval of approximately 3 weeks, are shown by the circle and triangle symbols. The locations of spherical plasmoids as a function of time are identified by the computed images at different times.
     The injected plasmoids are located along pathlines defined with  ($\theta_{\rm H}, \phi_{\rm H}$). 
      The  circles are cases of $\theta_{\rm H}=60^{\circ}$ and the  triangles are cases of 
     $\theta_{\rm H}=85^{\circ}$.
     $\phi_{\rm H}$ are chosen with a  $90^{\circ}$ difference.
     The trajectories are color coded according to their location relative to the 3D structure of the jet: according to the observer,  light-orange and orange indicate regions along the sides of the jet, while light-purple and purple correspond to the front and back regions. }
\label{fig:vapp_path}
\end{center}
\end{figure}

\begin{figure}
\centering
\includegraphics[width =0.45\textwidth]{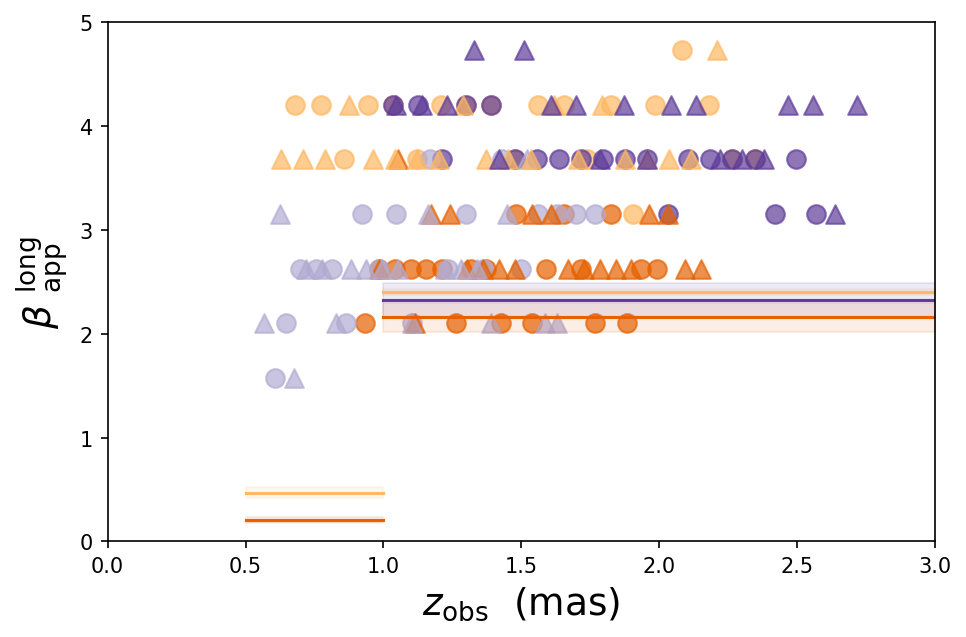}\\
\includegraphics[width =0.45\textwidth]
{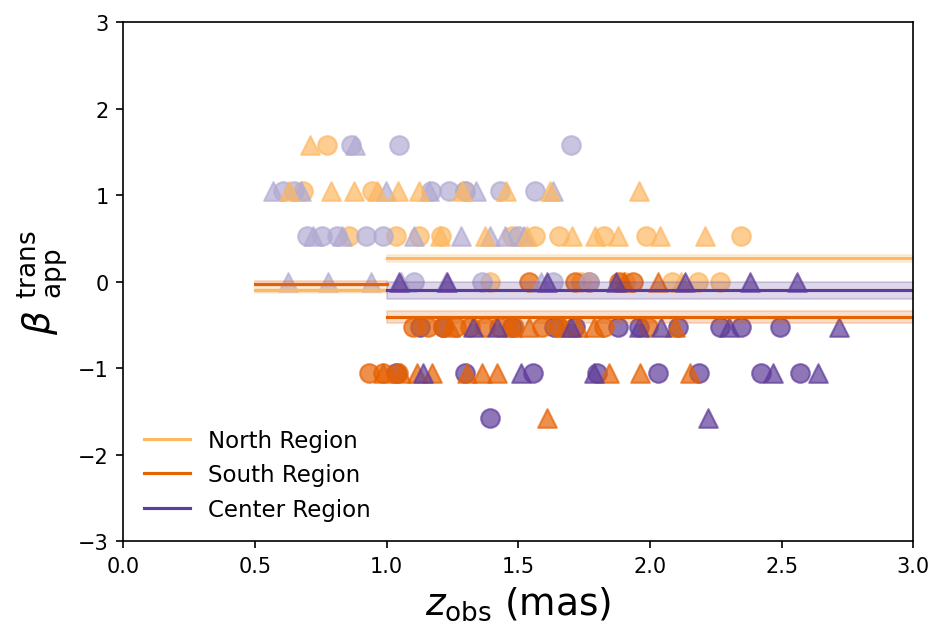}

    \caption{Apparent velocity derived from the trajectories shown in Figure \ref{fig:vapp_path}.  Circles corresponds to the case $\theta_{\rm H}=60^{\circ}$, triangles corresponds to the case $\theta_{\rm H}=85^{\circ}$. The symbols and colors follow the same format as in Figure \ref{fig:vapp_path}.  For reference,  the velocity field statistics reported by  \citet{mer2016} are overlapped with the horizontal lines and the corresponding shaded regions. See text for more details.} 
\label{fig:v_app}
\end{figure}

\begin{figure}
\centering
\includegraphics[width=0.5\textwidth]{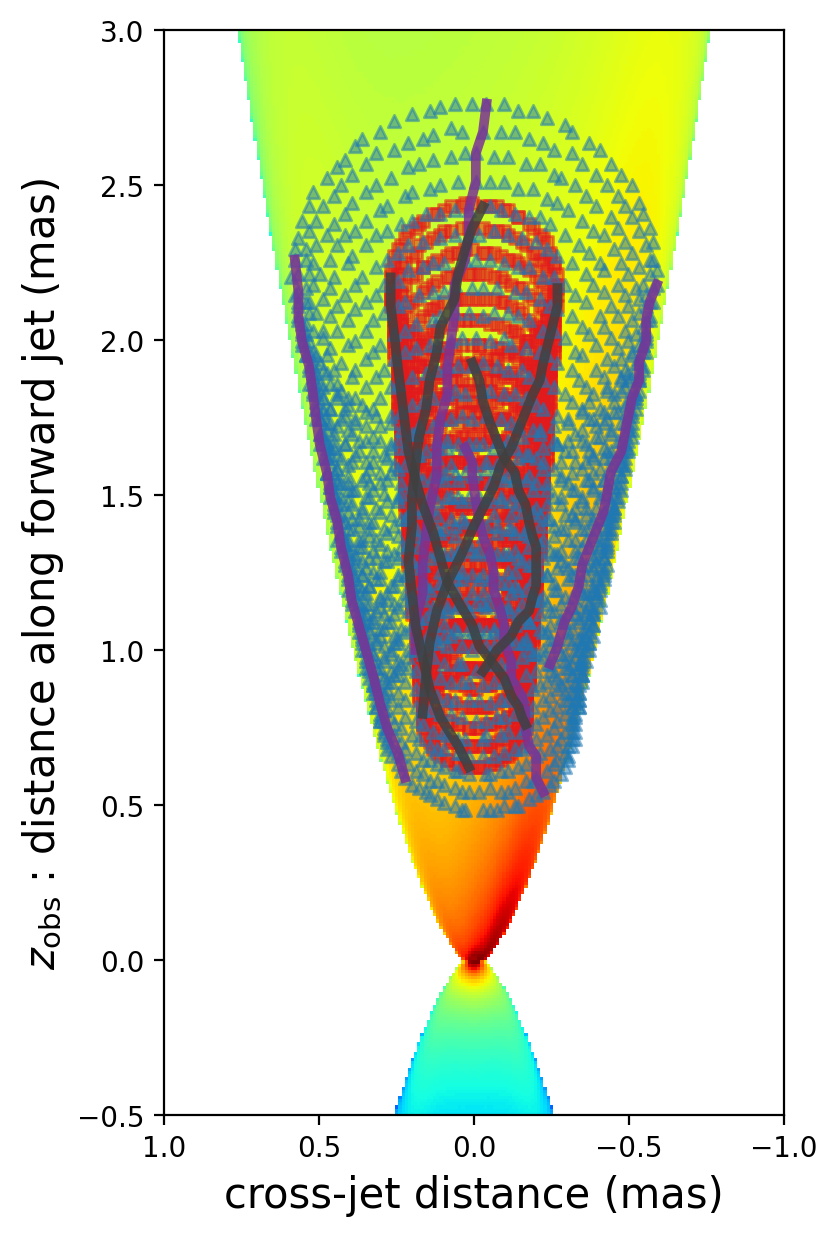}
        \caption{Ensemble of plasmoid trajectories corresponding to different streamline surfaces within the jet flow.
        The setup is the same as in Figure \ref{fig:vapp_path}, but for a family of $\phi_{\rm H}\in [0^{\circ}, 360^{\circ}]$ with a interval of $5^{\circ}$.
        The red squares are cases of $\theta_{\rm H}=30^{\circ}$, with representative trajectories in black. The blue triangles are cases of $\theta_{\rm H}=85^{\circ}$, with representative trajectories shown in purple (the same as the triangle trajectories shown in Figure \ref{fig:vapp_path}). }
\label{fig:vapp_all}
\end{figure}

\begin{figure}
\centering
\includegraphics[width=0.45\textwidth]{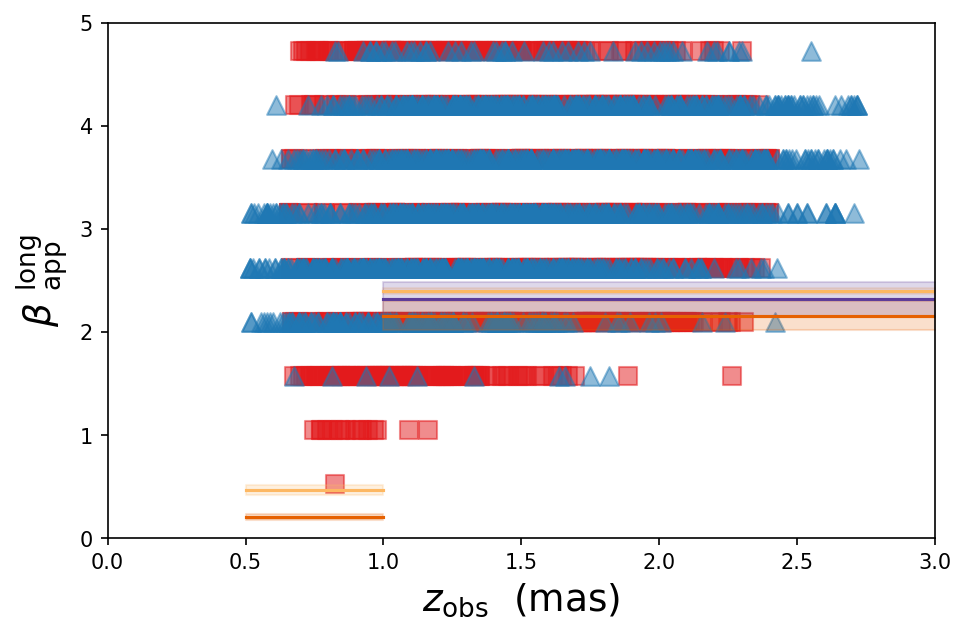} \hspace{0.05\textwidth}\\
\includegraphics[width=0.45\textwidth]
{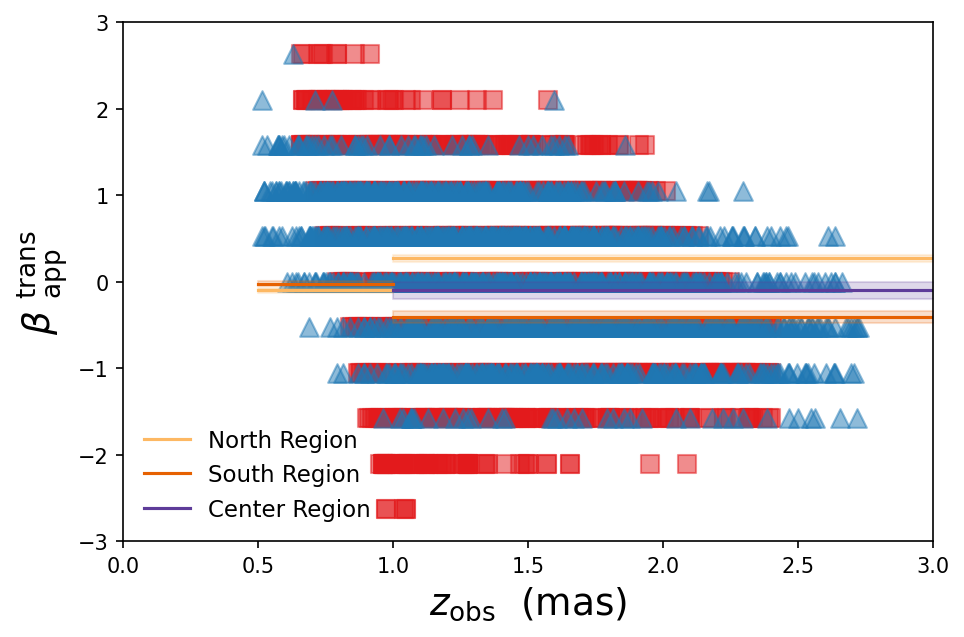} 

\caption{Same as Figure \ref{fig:v_app}, but for the apparent velocity derived from the trajectories shown in Figure \ref{fig:vapp_all}.  Red squares represent the case with $\theta_{\rm H} = 30^{\circ}$, while blue triangles correspond to $\theta_{\rm H} = 85^{\circ}$.}
\label{fig:vapp_all_dis}
\end{figure}

\subsection{Apparent Trajectories and Velocities of Injected Plasmoid}
To simulate the apparent motion of injected plasmoids within our jet model, we model a sequence of jet images that featured local spherical plasmoids with positions that varied along a selected pathline. The cadence between each image has 3-week intervals, extending over a duration of about 60 weeks. Noting that defining the location of the plasmoid in the blurred images can be biased by many factors, such as the beam size and the ratio between the background emission (see Figure \ref{fig:spot_sphere} and related discussion), in our procedure
we simply identify the location of the spherical plasmoids from the computed images. 
 
The trajectories are shown in Figure \ref{fig:vapp_path}. The distance along the observed forward jet axis is defined as $z_{\rm obs}$, and we set the black hole to be located at $z_{\rm obs}=0$.
The trajectories correspond to various pathlines as denoted in $\theta_{\rm H}=(60^{\circ}, 85 ^{\circ})$, represented by circles and triangles within the figure. The chosen values of $\phi_{\rm H}$ follow $\Delta \phi_{\rm H}=90^{\circ}$, and are color coded according to their distinct regions within the 3D jet structure.
It is obvious that the trajectories are poloidal, consistent with the analysis in \citep{mer2016}, since the pathline is dominated by the poloidal velocity (see also Figure \ref{fig:pathline}).

{It is expected that the observed plasmoid movement is influenced by multiple factors, including the properties of the injection (such as the location of the injection and the density or flux ratio relative to the background), the dynamics of the plasmoid itself, and the cooling timescale of the plasmoid. Although injection parameters are more uncertain due to microphysical processes, both jet motion and cooling timescale can be constrained by the MHD jet structure, as illustrated in Figure~\ref{fig:model_overview}.
Although the flow motion is faster near the jet boundary, as shown in Figure~\ref{fig:model_overview}(e), the synchrotron cooling timescale is actually longer near the jet spine, as illustrated in Figure~\ref{fig:model_overview}(c). The shorter cooling timescale near the jet edge may help explain why plasmoid motions are not observed to traverse the jet from one side to the other \citep[e.g.][]{mer2016}. Consequently, the complete trajectories shown in Figure \ref{fig:vapp_path} may not be entirely observable since the plasmoid undergoes cooling within the entire timeframe of Figure \ref{fig:vapp_path}, approximately 60 weeks. Given an estimated cooling timescale of $<$10 weeks, once injected, the plasmoid is expected to travel with a projected distance of $<$0.5 mas downstream, before it cools.}

The trajectories shown in Figure \ref{fig:vapp_path} exhibit superluminal motion. 
The apparent velocity $\beta=v/c$ can be decomposed into longitudinal $\beta_{\rm long}$ and transverse $\beta_{\rm trans}$ velocities along the forward jet axis $z_{\rm obs}$.
The apparent superluminal velocity of the bright spots is shown in Figure \ref{fig:v_app},  following the same color-coding used in Figure \ref{fig:vapp_path}.
 The computed apparent velocities have an error bar $\sim 0.525~c$ due to the limited image resolution and time cadence considered in Figure \ref{fig:vapp_path}.
It is obvious that $|\beta_{\rm long}| > |\beta_{\rm trans}|$ since the motion is mostly poloidal, and $\beta_{\rm trans}$ can be positive and negative depending on the transverse direction of the trajectories. 

By combining the WISE velocity field with the stacked cross-correlation analysis, \citet{mer2016} reported the average apparent velocity in the northern and southern regions of the jet across two distance ranges: $0.5$--$1$~mas and $1$--$4$~mas. Furthermore, the average velocity in the central region within the $1$--$4$~mas range was also reported \citep[Tables~2 and~3 of][]{mer2016}. We overlapped these reported values\footnote{The ``fast velocity component'' in Tables~2 and~3 of \citet[][]{mer2016}.}, as well as their statistically significant ranges, in Figure~\ref{fig:v_app} as horizontal lines and shaded regions. The color coding corresponds to the northern, southern, and central jet regions, consistent with the color scheme used in Figure~\ref{fig:vapp_path}. 

Due to the difficulty of accurately identifying its location in the presence of background emission and the finite observational beam size,  as illustrated in Figure~\ref{fig:spot_sphere},  the apparent velocity of the observed plasmoid can generally appear to be lower than the values modeled in Figure~\ref{fig:v_app}, which is calculated directly in the model image plane.  Therefore, the modeled scattered distribution of apparent velocities is sufficient to explain the inferred velocities of the M87 jet reported in \citet[][their Figure~3]{mer2016}.
In addition, it is observed that in the lower panel of Figure~\ref{fig:v_app}, the distribution of plasmoids on opposing sides of the jet manifests opposite signs of apparent transverse velocity $\beta^{\rm trans}_{\rm app}$. More precisely, the light-orange (or orange) markers in Figure~\ref{fig:vapp_path} correspond to $\beta^{\rm trans}_{\rm app}>0$ (or $\beta^{\rm trans}_{\rm app}<0$), aligned with the light-orange line (or orange line) within the $1$--$4$~mas range, where the observed plasmoid motions are anticipated to be less influenced by intricate background emissions.

To further compare the motion of the plamoid on fixed streamline surfaces (defined by  $\Psi$=constant) near the jet spine and the jet boundary, Figures \ref{fig:vapp_all} and Figure \ref{fig:vapp_all_dis} display the path and apparent velocity distribution of the plasmoid injected at different initial azimuthal locations along the streamline surface with $\theta_{\rm H}=30^{\circ}$ (red squares) and $\theta_{\rm H}=85^{\circ}$ (blue triangles). The collective locations of the plasmoid at different times clearly indicate the separated streamline surface.
The black and purple lines in Figure \ref{fig:vapp_all} respectively show representative trajectories for plasmoids along the streamline surface closer to the jet spine
and those closer to the jet boundary.
The trajectory along the jet spine follows a more twisted and curved path, in contrast to the relatively straighter motion near the jet boundary. This results in a diminished value for $\beta^{\rm long}_{\rm app}$ in the smaller $z_{\rm obs}$ (as observed in the upper panel of Figure \ref{fig:vapp_all_dis}) and a wider and more dispersed distribution of $\beta^{\rm trans}_{\rm app}$ (as evidenced in the lower panel of Figure \ref{fig:vapp_all_dis}).

\section{Summary and Implications}\label{sec:summary}
In this work, by modeling  observed M87 jet images at 43 GHz, we investigate the influence of asymmetric plasmoids injected within the M87 jet on subparsec jet properties, including apparent trajectories, velocity, and the resulting alteration of jet morphology. 
Our jet model extends the steady, axisymmetric force-free jet model introduced in \citet[][] {broderick2009imaging} and \citet{takahashi2018fast}, with two key modifications motivated by observations. First, we consider that the jet boundary is defined by the large-scale magnetic field attaching to the black hole event horizon on the equatorial plane. This modification is consistent with the observed jet width (Figure \ref{fig:jet_image_bg}).
Second,  we calibrate the terminal velocity of the jet velocity to align with observation, incorporating modifications  originally suggested in \citet{lu2014}. 
Our model considers both a steady, axisymmetric plasma mass loading and an intermittently injected, asymmetric plasmoid within the jet. 

Inspired by the comprehensive analysis of the morphologies and kinematics of M87 at 43 GHz \citep{mer2016,walker2018structure}, we concentrate on the same frequency to compare our model jet images with these analyses. Our model with steady axisymmetric mass loading (Figure \ref{fig:jet_image_bg}) demonstrates limb-brightened features that visually resemble the overall 43 GHz observations \citep{walker2018structure}. We proceed to demonstrate how the model's jet morphologies are influenced by asymmetric plasmoids within the jet. The resulting jet morphologies depend both on the selected beam size and on the intrinsic emissivity of the plasmoids (see Figure \ref{fig:spot_sphere}). Furthermore, the impact of shearing plasmoids is examined, particularly when these exhibit an elongated shape along the pathline (see Figures \ref{fig:asym_spot_left} and \ref{fig:spot_all_around}). It is demonstrated that the presence of an asymmetric plasmoid within the jet can lead to significant diversity in jet morphologies, but only a limited variation in flux (e.g. $\Delta F<0.1$ Jy). For example, the brightness of the jet limbs may vary over time, offering a natural explanation for the temporal variation in the brightness of the M87 jet limbs at subparsec scales \citep{walker2018structure}. In practice, there may be multiple injections of plasmoids,  each exhibiting a shearing distance determined by the cooling timescale.

Based on the flow velocity in our model, the apparent velocity of the asymmetric plasmoids injected within the jet can adequately account for the observed trajectories and kinematics of the jets at the parsec scales \citet{mer2016}. The model trajectories of the plasmoid are predominantly poloidal, attributed to the predominance of poloidal velocity in regions significantly larger than the light cylinder (see Figures \ref{fig:vapp_path} --\ref{fig:vapp_all_dis}). 
In addition, the relative cooling timescales between the jet spine and boundary regions constrain the lifetimes of plasmoids injected at different locations within the jet.
We emphasize that asymmetric plasmoid injections, whose motion and morphology are governed by poloidal velocity and radiative cooling, play a crucial role in determining the subparsec-scale observable properties of the M87 jet.
The development of more advanced models concerning the dynamical evolution of jet emissions would be achievable through a deeper understanding of the mechanisms responsible for the injection of high-energy particles.

\begin{acknowledgments}

The authors thank Keiichi Asada and Britt Jeter for their invaluable discussions.  The author also thanks the anonymous referee for constructive comments, which significantly improved the quality of the paper. H.Y.P. acknowledges the support received from the Yushan Young Scholar Program under the Ministry of Education (MoE) of Taiwan, as well as the funding provided by the National Science and Technology Council (NSTC) of Taiwan through grant 112-2112-M-003-010-MY3. This research has benefited from the resources of NASA's Astrophysics Data System.
\end{acknowledgments}

\appendix
\section{Images at 15, 22, and 86 frequencies}\label{app:multifreq_img}

In this Appendix, we present both the background jet images and example jet images with shearing plasmoids at frequencies other than 43 GHz. The shearing spot parameters adopted are the same as those used in Figure \ref{fig:asym_spot_left}(f), namely,
$
(dr_{\rm inj},~dj_{\rm inj},~dl_{\rm inj}) = \left(15~r_{g},~20,~600~r_{g}\right)
$.

\begin{figure*}[ht]
    \centering
      \includegraphics[width =0.45\textwidth]{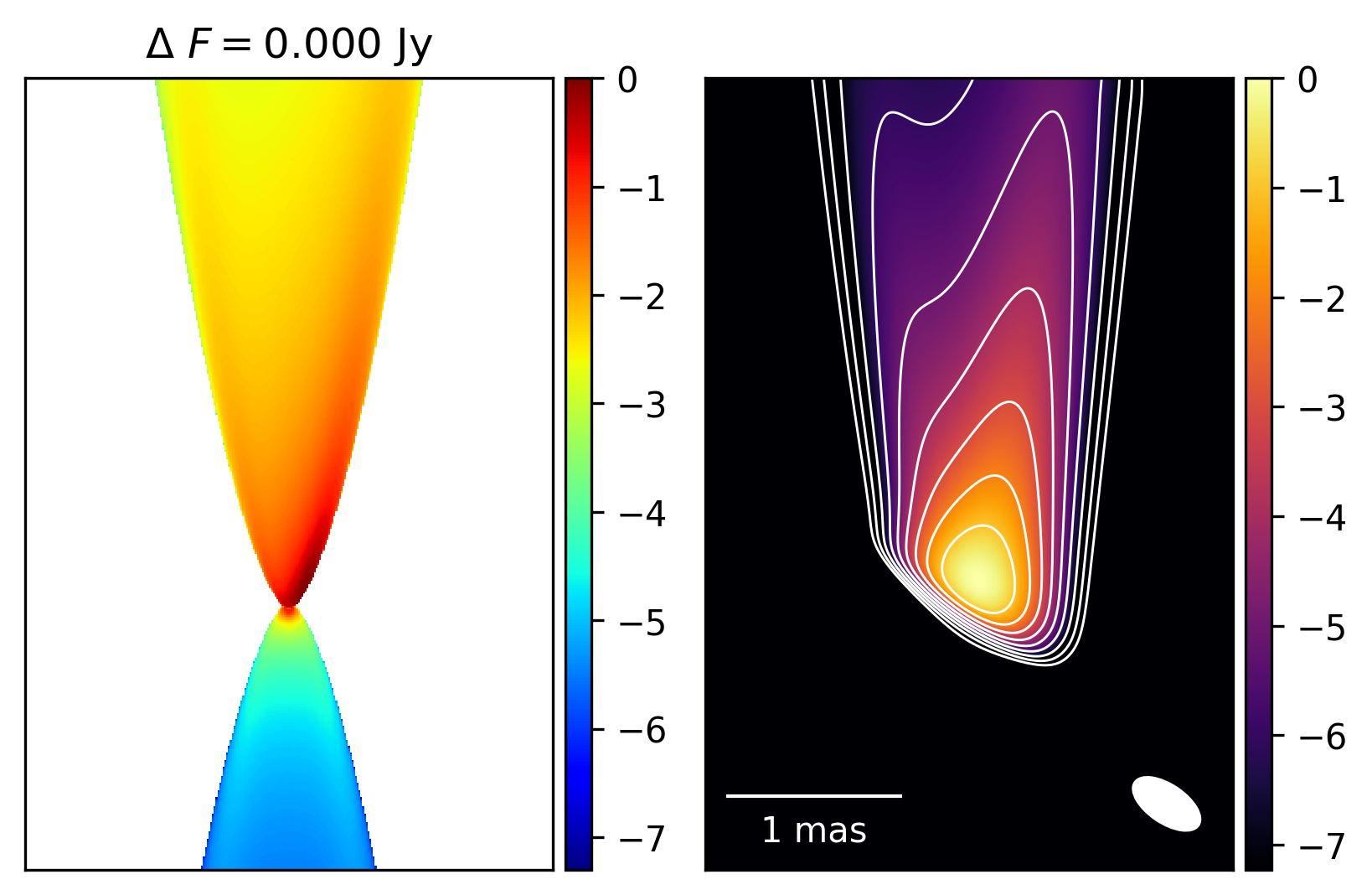}
    \includegraphics[width =0.45\textwidth]{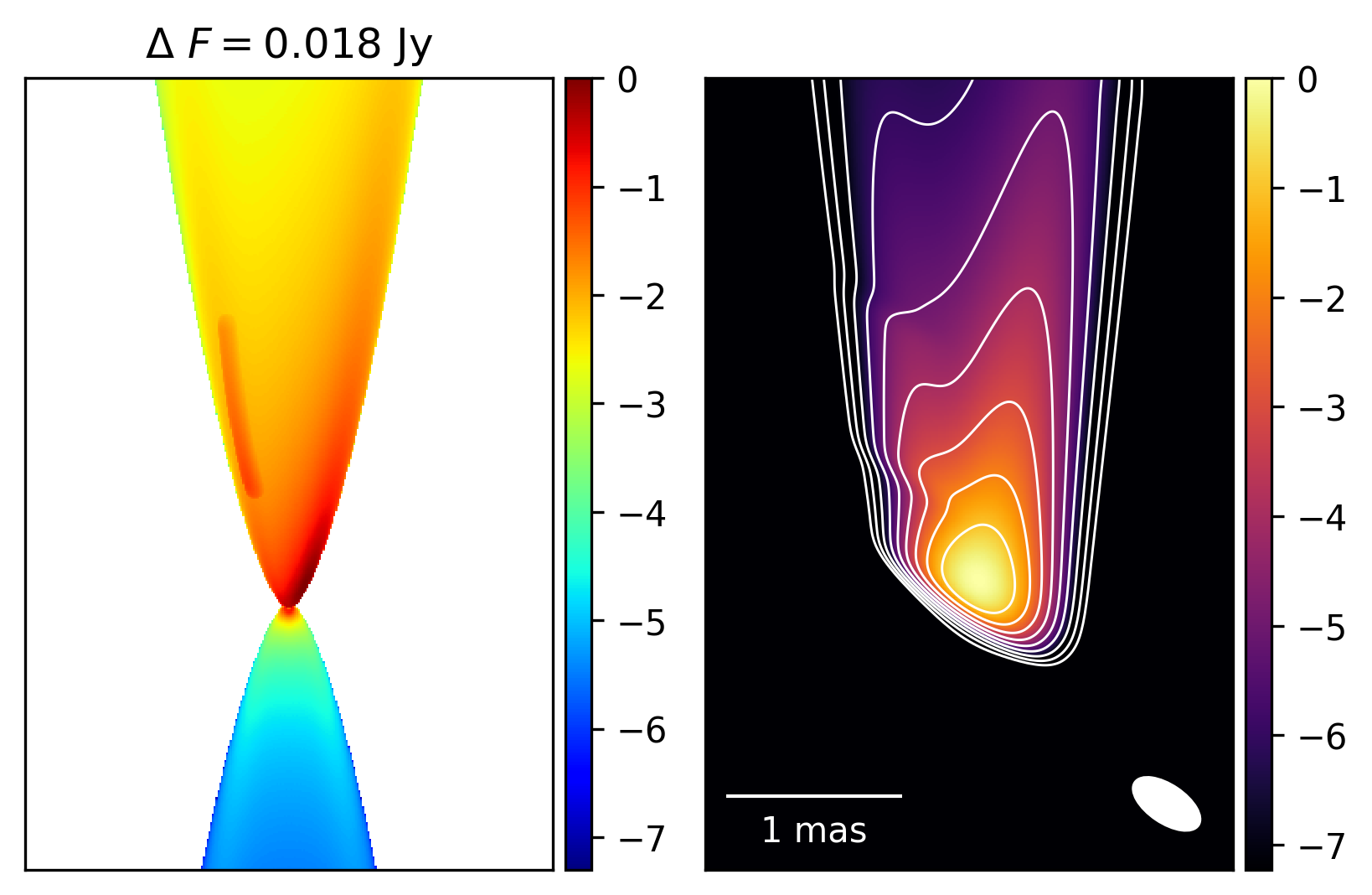}
    \caption{Model images at 15 GHz, without ($\Delta F=0$ Jy) and with ($\Delta F>0$ Jy) asymmetric shearing plasmoid. The shearing plasmoids have the same parameters as those used for Figure \ref{fig:asym_spot_left}(f).
     }
\label{fig:spot_15ghz}
\end{figure*}

\begin{figure*}[ht]
    \centering
      \includegraphics[width =0.45\textwidth]{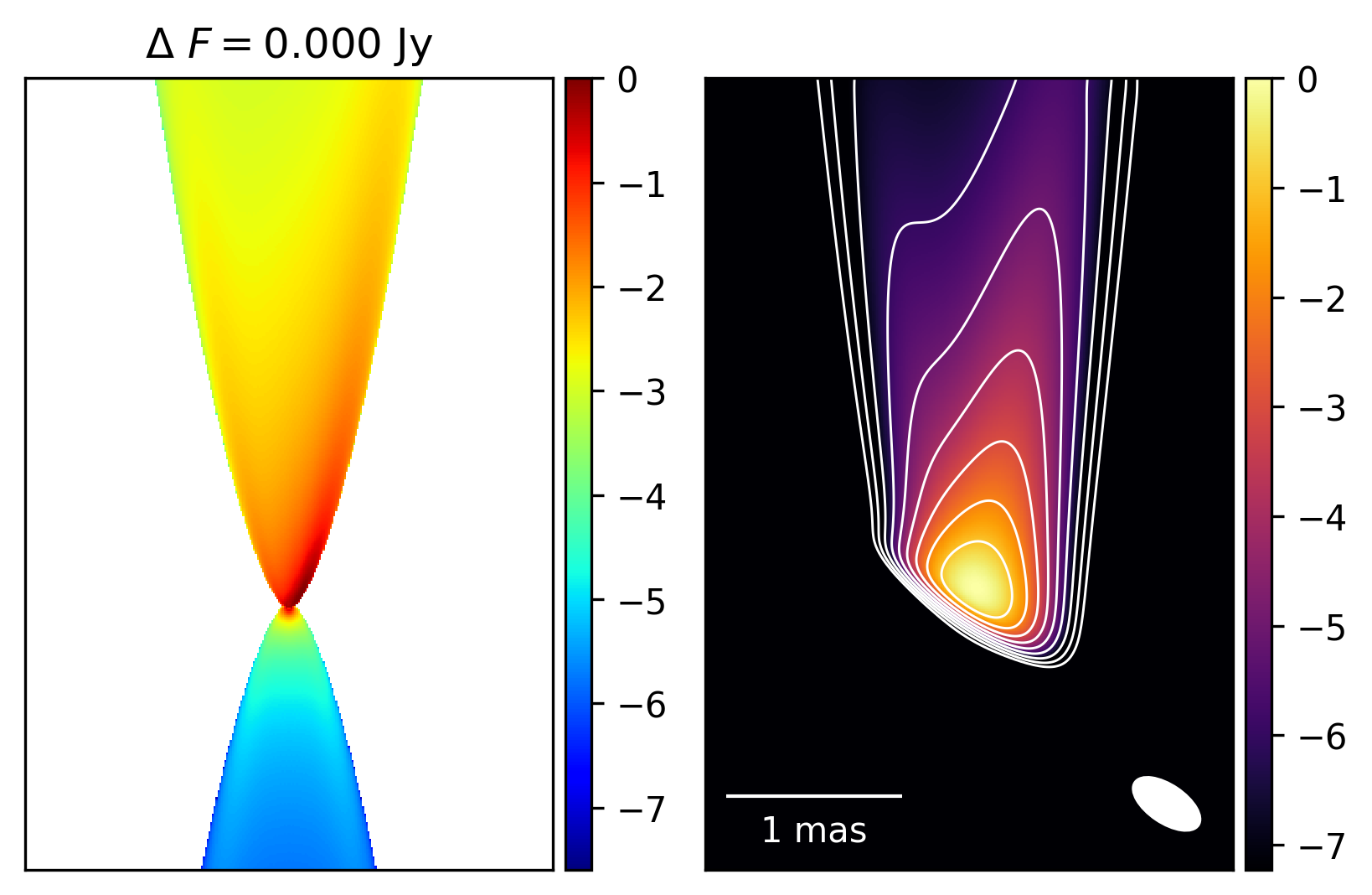}
    \includegraphics[width =0.45\textwidth]{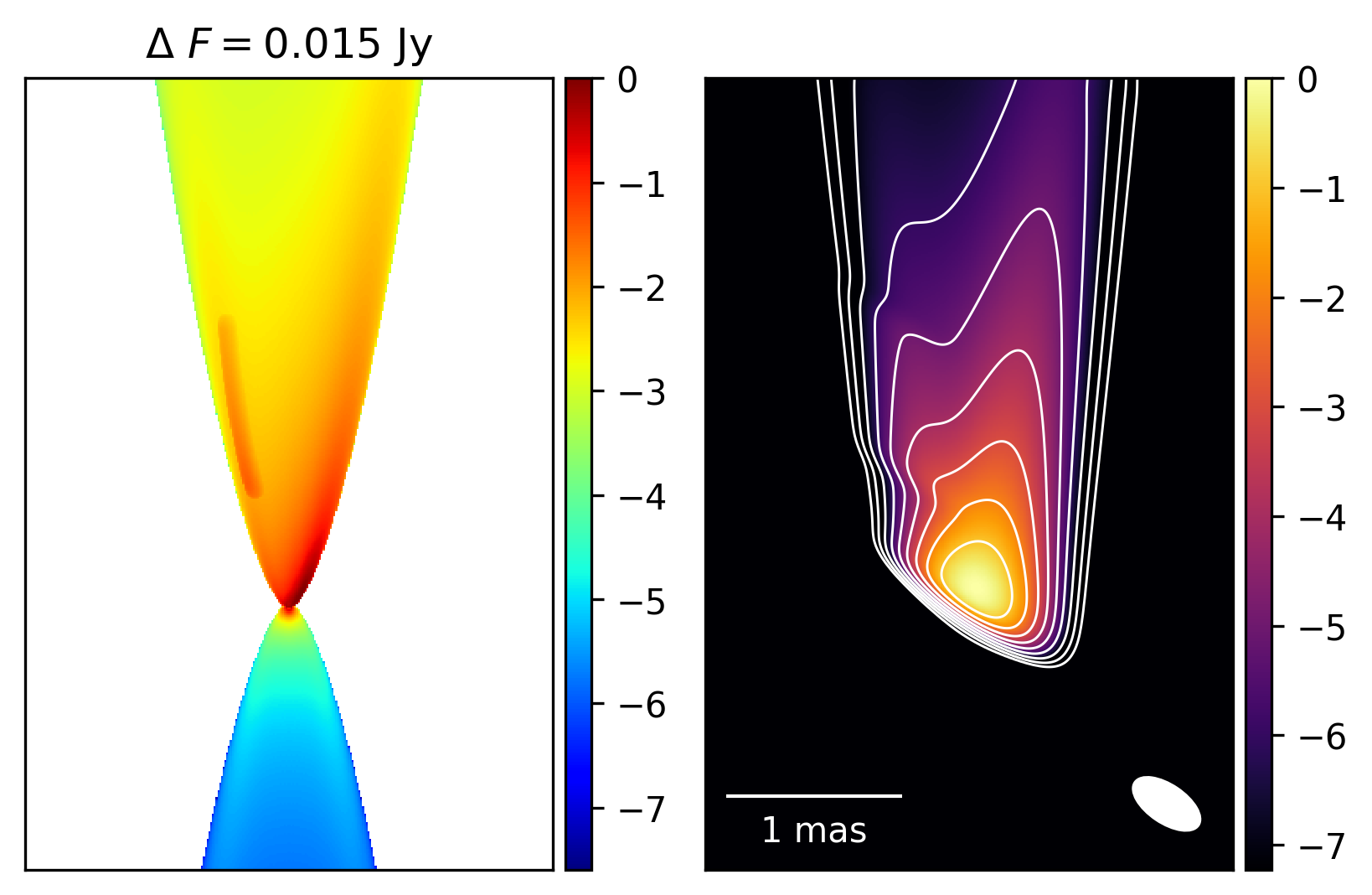}
    \caption{  Same as Figure \ref{fig:spot_15ghz}, but at 22 GHz.
     }
\label{fig:spot_22ghz}
\end{figure*}

\begin{figure*}[ht]
    \centering
      
      \includegraphics[width =0.45\textwidth]{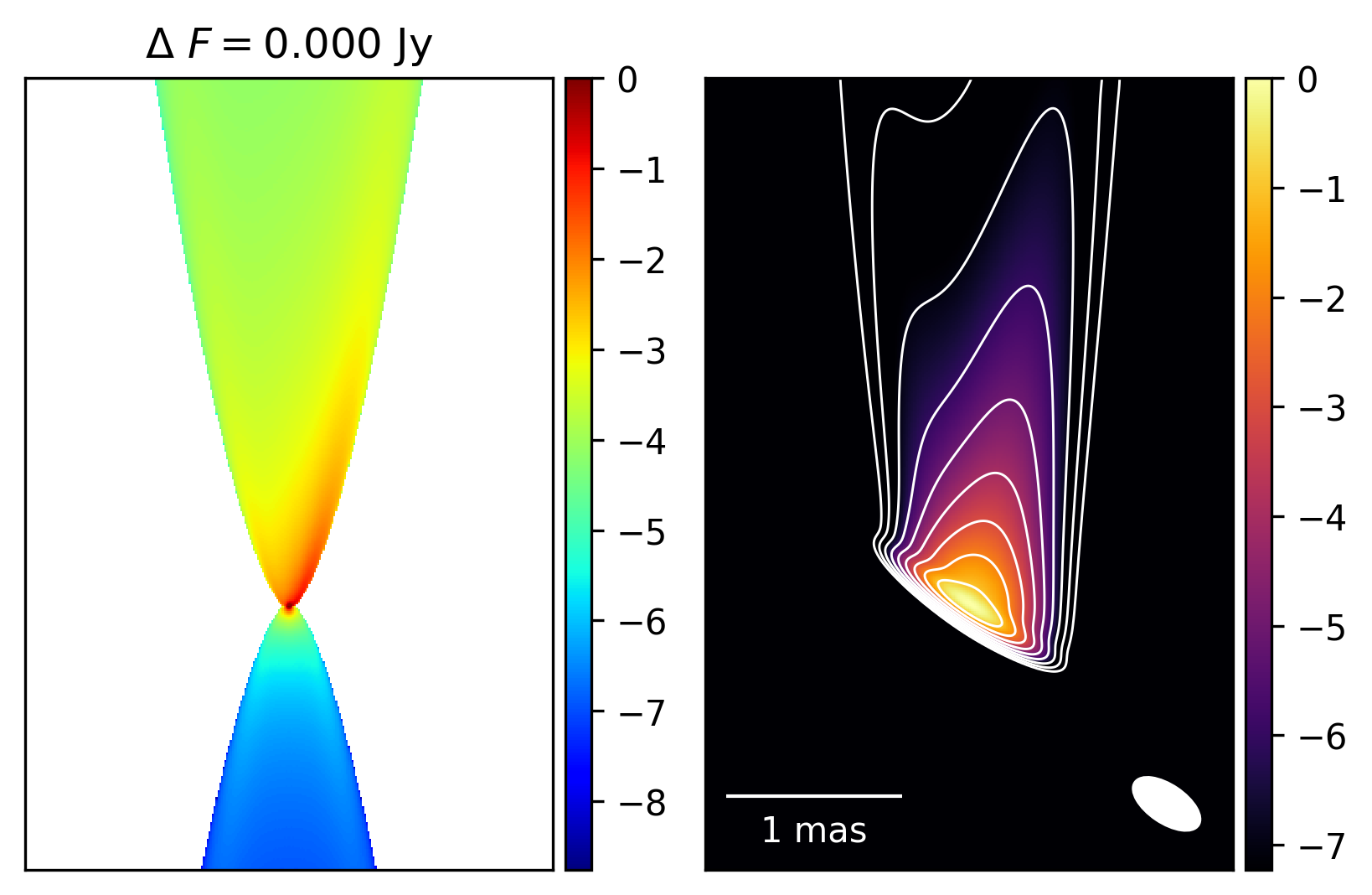}
    \includegraphics[width =0.45\textwidth]{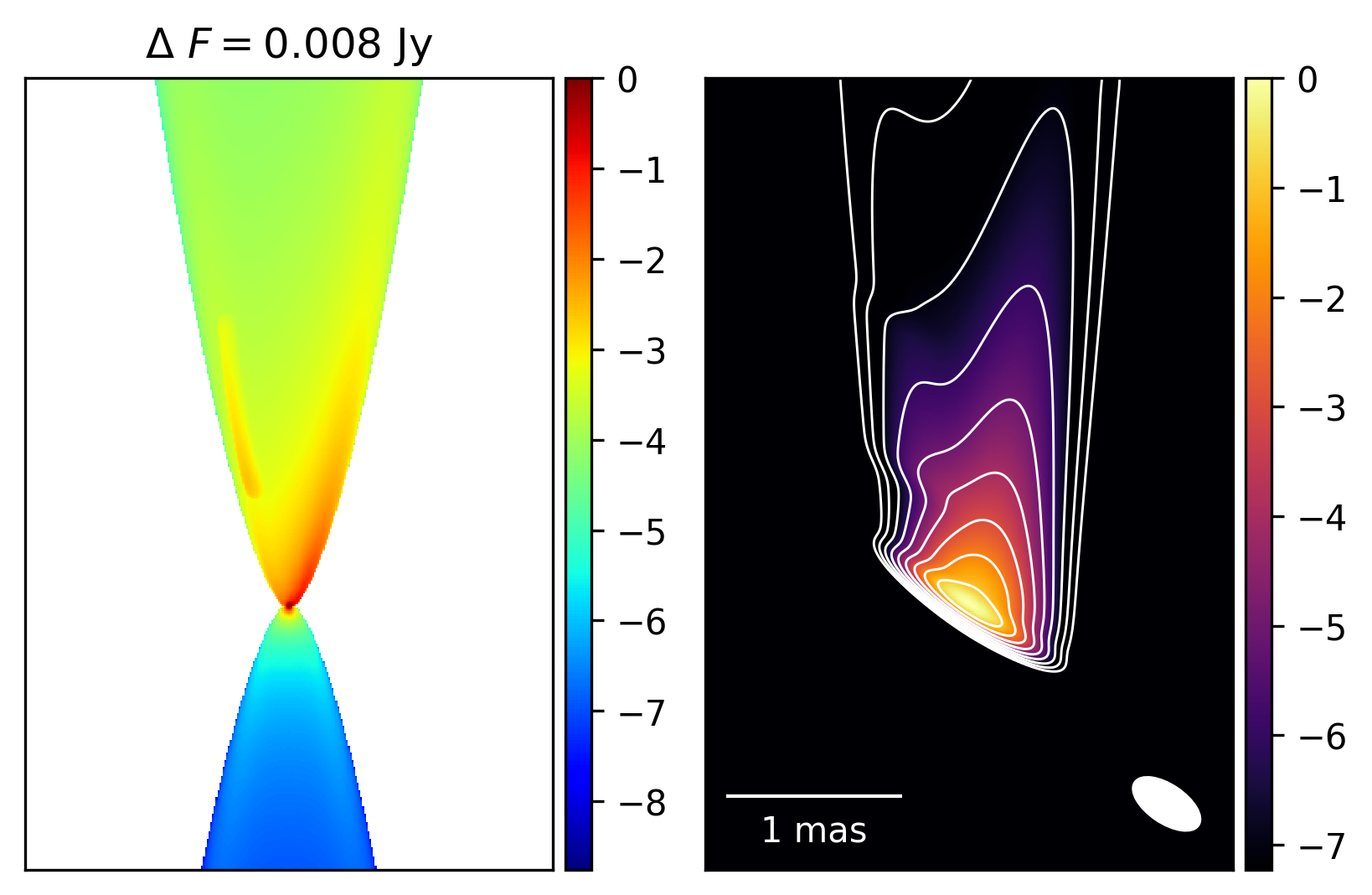}\\
    \caption{  Same as Figure \ref{fig:spot_15ghz}, but at 86GHz.
     }
\label{fig:spot_86ghz}
\end{figure*}

Figures \ref{fig:spot_15ghz} to \ref{fig:spot_86ghz} present the model images at frequencies of 15, 22, and 86 GHz, respectively. For demonstration purposes, the beam size employed for the blurred images at 43 GHz is also applied to the blurred images at these frequencies. As anticipated, at higher frequencies, the jet emission is less extended, and the emission core shifts toward the black hole due to the optical depth effect, commonly referred to as the "core-shift" effect \citep[e.g.][]{hada2011}. Similarly to the case at 43 GHz shown in Figure \ref{fig:asym_spot_left}(f), it is shown that asymmetric plasmoid loading can lead to a visually discernible alteration in the morphology of the jet, with a minor modification in the total flux $\Delta F$.


\bibliography{references}{}

\begin{thebibliography}{}
\expandafter\ifx\csname natexlab\endcsname\relax\def\natexlab#1{#1}\fi
\providecommand{\url}[1]{\href{#1}{#1}}
\providecommand{\dodoi}[1]{doi:~\href{http://doi.org/#1}{\nolinkurl{#1}}}
\providecommand{\doeprint}[1]{\href{http://ascl.net/#1}{\nolinkurl{http://ascl.net/#1}}}
\providecommand{\doarXiv}[1]{\href{https://arxiv.org/abs/#1}{\nolinkurl{https://arxiv.org/abs/#1}}}

\bibitem[{K. {Asada} \& M. {Nakamura}(2012){Asada} \& {Nakamura}}]{asada2012}
{Asada}, K., \& {Nakamura}, M. 2012, \bibinfo{title}{{The Structure of the M87 Jet: A Transition from Parabolic to Conical Streamlines},} \apjl, 745, L28, \dodoi{10.1088/2041-8205/745/2/L28}

\bibitem[{V.~S. Beskin(2009)Beskin}]{beskin2009mhd}
Beskin, V.~S. 2009, MHD flows in compact astrophysical objects: accretion, winds and jets (Springer Science \& Business Media)

\bibitem[{R.~D. Blandford \& R.~L. Znajek(1977)Blandford \& Znajek}]{blandford1977electromagnetic}
Blandford, R.~D., \& Znajek, R.~L. 1977, \bibinfo{title}{Electromagnetic extraction of energy from Kerr black holes,} Monthly Notices of the Royal Astronomical Society, 179, 433

\bibitem[{A.~E. Broderick \& A. Loeb(2009)Broderick \& Loeb}]{broderick2009imaging}
Broderick, A.~E., \& Loeb, A. 2009, \bibinfo{title}{Imaging the black hole silhouette of M87: implications for jet formation and black hole spin,} The Astrophysical Journal, 697, 1164

\bibitem[{A. {Chael} {et~al.}(2019){Chael}, {Narayan}, \& {Johnson}}]{chael2019}
{Chael}, A., {Narayan}, R., \& {Johnson}, M.~D. 2019, \bibinfo{title}{{Two-temperature, Magnetically Arrested Disc simulations of the jet from the supermassive black hole in M87},} \mnras, 486, 2873, \dodoi{10.1093/mnras/stz988}

\bibitem[{A. {Cruz-Osorio} {et~al.}(2022){Cruz-Osorio}, {Fromm}, {Mizuno}, {Nathanail}, {Younsi}, {Porth}, {Davelaar}, {Falcke}, {Kramer}, \& {Rezzolla}}]{cruz2022}
{Cruz-Osorio}, A., {Fromm}, C.~M., {Mizuno}, Y., {et~al.} 2022, \bibinfo{title}{{State-of-the-art energetic and morphological modelling of the launching site of the M87 jet},} Nature Astronomy, 6, 103, \dodoi{10.1038/s41550-021-01506-w}

\bibitem[{J. {Davelaar} {et~al.}(2019){Davelaar}, {Olivares}, {Porth}, {Bronzwaer}, {Janssen}, {Roelofs}, {Mizuno}, {Fromm}, {Falcke}, \& {Rezzolla}}]{jordy2019}
{Davelaar}, J., {Olivares}, H., {Porth}, O., {et~al.} 2019, \bibinfo{title}{{Modeling non-thermal emission from the jet-launching region of M 87 with adaptive mesh refinement},} \aap, 632, A2, \dodoi{10.1051/0004-6361/201936150}

\bibitem[{ {Event Horizon Telescope Collaboration} {et~al.}(2019{\natexlab{a}}){Event Horizon Telescope Collaboration}, {Akiyama}, {Alberdi}, {Alef}, {Asada}, {Azulay}, {Baczko}, {Ball}, {Balokovi{\'c}}, {Barrett}, {Bintley}, {Blackburn}, {Boland}, {Bouman}, {Bower}, {Bremer}, {Brinkerink}, {Brissenden}, {Britzen}, {Broderick}, {Broguiere}, {Bronzwaer}, {Byun}, {Carlstrom}, {Chael}, {Chan}, {Chatterjee}, {Chatterjee}, {Chen}, {Chen}, {Cho}, {Christian}, {Conway}, {Cordes}, {Crew}, {Cui}, {Davelaar}, {De Laurentis}, {Deane}, {Dempsey}, {Desvignes}, {Dexter}, {Doeleman}, {Eatough}, {Falcke}, {Fish}, {Fomalont}, {Fraga-Encinas}, {Freeman}, {Friberg}, {Fromm}, {G{\'o}mez}, {Galison}, {Gammie}, {Garc{\'\i}a}, {Gentaz}, {Georgiev}, {Goddi}, {Gold}, {Gu}, {Gurwell}, {Hada}, {Hecht}, {Hesper}, {Ho}, {Ho}, {Honma}, {Huang}, {Huang}, {Hughes}, {Ikeda}, {Inoue}, {Issaoun}, {James}, {Jannuzi}, {Janssen}, {Jeter}, {Jiang}, {Johnson}, {Jorstad}, {Jung}, {Karami}, {Karuppusamy}, {Kawashima}, {Keating}, {Kettenis}, {Kim},
  {Kim}, {Kim}, {Kino}, {Koay}, {Koch}, {Koyama}, {Kramer}, {Kramer}, {Krichbaum}, {Kuo}, {Lauer}, {Lee}, {Li}, {Li}, {Lindqvist}, {Liu}, {Liuzzo}, {Lo}, {Lobanov}, {Loinard}, {Lonsdale}, {Lu}, {MacDonald}, {Mao}, {Markoff}, {Marrone}, {Marscher}, {Mart{\'\i}-Vidal}, {Matsushita}, {Matthews}, {Medeiros}, {Menten}, {Mizuno}, {Mizuno}, {Moran}, {Moriyama}, {Moscibrodzka}, {M{\"u}ller}, {Nagai}, {Nagar}, {Nakamura}, {Narayan}, {Narayanan}, {Natarajan}, {Neri}, {Ni}, {Noutsos}, {Okino}, {Olivares}, {Ortiz-Le{\'o}n}, {Oyama}, {{\"O}zel}, {Palumbo}, {Patel}, {Pen}, {Pesce}, {Pi{\'e}tu}, {Plambeck}, {PopStefanija}, {Porth}, {Prather}, {Preciado-L{\'o}pez}, {Psaltis}, {Pu}, {Ramakrishnan}, {Rao}, {Rawlings}, {Raymond}, {Rezzolla}, {Ripperda}, {Roelofs}, {Rogers}, {Ros}, {Rose}, {Roshanineshat}, {Rottmann}, {Roy}, {Ruszczyk}, {Ryan}, {Rygl}, {S{\'a}nchez}, {S{\'a}nchez-Arguelles}, {Sasada}, {Savolainen}, {Schloerb}, {Schuster}, {Shao}, {Shen}, {Small}, {Sohn}, {SooHoo}, {Tazaki}, {Tiede}, {Tilanus}, {Titus}, {Toma},
  {Torne}, {Trent}, {Trippe}, {Tsuda}, {van Bemmel}, {van Langevelde}, {van Rossum}, {Wagner}, {Wardle}, {Weintroub}, {Wex}, {Wharton}, {Wielgus}, {Wong}, {Wu}, {Young}, \& {Young}}]{eht2019a}
{Event Horizon Telescope Collaboration}, {Akiyama}, K., {Alberdi}, A., {et~al.} 2019{\natexlab{a}}, \bibinfo{title}{{First M87 Event Horizon Telescope Results. I. The Shadow of the Supermassive Black Hole},} \apjl, 875, L1, \dodoi{10.3847/2041-8213/ab0ec7}

\bibitem[{ {Event Horizon Telescope Collaboration} {et~al.}(2019{\natexlab{b}}){Event Horizon Telescope Collaboration}, {Akiyama}, {Alberdi}, {Alef}, {Asada}, {Azulay}, {Baczko}, {Ball}, {Balokovi{\'c}}, {Barrett}, {Bintley}, {Blackburn}, {Boland}, {Bouman}, {Bower}, {Bremer}, {Brinkerink}, {Brissenden}, {Britzen}, {Broderick}, {Broguiere}, {Bronzwaer}, {Byun}, {Carlstrom}, {Chael}, {Chan}, {Chatterjee}, {Chatterjee}, {Chen}, {Chen}, {Cho}, {Christian}, {Conway}, {Cordes}, {Crew}, {Cui}, {Davelaar}, {De Laurentis}, {Deane}, {Dempsey}, {Desvignes}, {Dexter}, {Doeleman}, {Eatough}, {Falcke}, {Fish}, {Fomalont}, {Fraga-Encinas}, {Friberg}, {Fromm}, {G{\'o}mez}, {Galison}, {Gammie}, {Garc{\'\i}a}, {Gentaz}, {Georgiev}, {Goddi}, {Gold}, {Gu}, {Gurwell}, {Hada}, {Hecht}, {Hesper}, {Ho}, {Ho}, {Honma}, {Huang}, {Huang}, {Hughes}, {Ikeda}, {Inoue}, {Issaoun}, {James}, {Jannuzi}, {Janssen}, {Jeter}, {Jiang}, {Johnson}, {Jorstad}, {Jung}, {Karami}, {Karuppusamy}, {Kawashima}, {Keating}, {Kettenis}, {Kim}, {Kim}, {Kim},
  {Kino}, {Koay}, {Koch}, {Koyama}, {Kramer}, {Kramer}, {Krichbaum}, {Kuo}, {Lauer}, {Lee}, {Li}, {Li}, {Lindqvist}, {Liu}, {Liuzzo}, {Lo}, {Lobanov}, {Loinard}, {Lonsdale}, {Lu}, {MacDonald}, {Mao}, {Markoff}, {Marrone}, {Marscher}, {Mart{\'\i}-Vidal}, {Matsushita}, {Matthews}, {Medeiros}, {Menten}, {Mizuno}, {Mizuno}, {Moran}, {Moriyama}, {Moscibrodzka}, {Mul{\ensuremath{\ddot{}}}ler}, {Nagai}, {Nagar}, {Nakamura}, {Narayan}, {Narayanan}, {Natarajan}, {Neri}, {Ni}, {Noutsos}, {Okino}, {Olivares}, {Oyama}, {{\"O}zel}, {Palumbo}, {Patel}, {Pen}, {Pesce}, {Pi{\'e}tu}, {Plambeck}, {PopStefanija}, {Porth}, {Prather}, {Preciado-L{\'o}pez}, {Psaltis}, {Pu}, {Ramakrishnan}, {Rao}, {Rawlings}, {Raymond}, {Rezzolla}, {Ripperda}, {Roelofs}, {Rogers}, {Ros}, {Rose}, {Roshanineshat}, {Rottmann}, {Roy}, {Ruszczyk}, {Ryan}, {Rygl}, {S{\'a}nchez}, {S{\'a}nchez-Arguelles}, {Sasada}, {Savolainen}, {Schloerb}, {Schuster}, {Shao}, {Shen}, {Small}, {Sohn}, {SooHoo}, {Tazaki}, {Tiede}, {Tilanus}, {Titus}, {Toma}, {Torne},
  {Trent}, {Trippe}, {Tsuda}, {van Bemmel}, {van Langevelde}, {van Rossum}, {Wagner}, {Wardle}, {Weintroub}, {Wex}, {Wharton}, {Wielgus}, {Wong}, {Wu}, {Young}, {Young}, {Younsi}, {Yuan}, {Yuan}, {Zensus}, {Zhao}, {Zhao}, {Zhu}, {Anczarski}, {Baganoff}, {Eckart}, {Farah}, {Haggard}, {Meyer-Zhao}, {Michalik}, {Nadolski}, {Neilsen}, {Nishioka}, {Nowak}, {Pradel}, {Primiani}, {Souccar}, {Vertatschitsch}, {Yamaguchi}, \& {Zhang}}]{2019ehtV}
{Event Horizon Telescope Collaboration}, {Akiyama}, K., {Alberdi}, A., {et~al.} 2019{\natexlab{b}}, \bibinfo{title}{{First M87 Event Horizon Telescope Results. V. Physical Origin of the Asymmetric Ring},} \apjl, 875, L5, \dodoi{10.3847/2041-8213/ab0f43}

\bibitem[{C.~M. {Fromm} {et~al.}(2022){Fromm}, {Cruz-Osorio}, {Mizuno}, {Nathanail}, {Younsi}, {Porth}, {Olivares}, {Davelaar}, {Falcke}, {Kramer}, \& {Rezzolla}}]{fromm2022}
{Fromm}, C.~M., {Cruz-Osorio}, A., {Mizuno}, Y., {et~al.} 2022, \bibinfo{title}{{Impact of non-thermal particles on the spectral and structural properties of M87},} \aap, 660, A107, \dodoi{10.1051/0004-6361/202142295}

\bibitem[{K. {Hada} {et~al.}(2011){Hada}, {Doi}, {Kino}, {Nagai}, {Hagiwara}, \& {Kawaguchi}}]{hada2011}
{Hada}, K., {Doi}, A., {Kino}, M., {et~al.} 2011, \bibinfo{title}{{An origin of the radio jet in M87 at the location of the central black hole},} \nat, 477, 185, \dodoi{10.1038/nature10387}

\bibitem[{K. Hada {et~al.}(2013)Hada, Kino, Doi, Nagai, Honma, Hagiwara, Giroletti, Giovannini, \& Kawaguchi}]{hada2013innermost}
Hada, K., Kino, M., Doi, A., {et~al.} 2013, \bibinfo{title}{The innermost collimation structure of the M87 jet down to~ 10 Schwarzschild radii,} The Astrophysical Journal, 775, 70

\bibitem[{K. Hada {et~al.}(2016)Hada, Kino, Doi, Nagai, Honma, Akiyama, Tazaki, Lico, Giroletti, Giovannini, {et~al.}}]{hada2016high}
Hada, K., Kino, M., Doi, A., {et~al.} 2016, \bibinfo{title}{High-sensitivity 86 GHz (3.5 mm) VLBI observations of M87: deep imaging of the jet base at a resolution of 10 Schwarzschild radii,} The Astrophysical Journal, 817, 131

\bibitem[{S. Hirose {et~al.}(2004)Hirose, Krolik, De~Villiers, \& Hawley}]{hirose2004magnetically}
Hirose, S., Krolik, J.~H., De~Villiers, J.-P., \& Hawley, J.~F. 2004, \bibinfo{title}{Magnetically driven accretion flows in the Kerr metric. II. Structure of the magnetic field,} The Astrophysical Journal, 606, 1083

\bibitem[{B. {Jeter} {et~al.}(2020){Jeter}, {Broderick}, \& {Gold}}]{jeter2020}
{Jeter}, B., {Broderick}, A.~E., \& {Gold}, R. 2020, \bibinfo{title}{{Differentiating disc and black hole-driven jets with EHT images of variability in M87},} \mnras, 493, 5606, \dodoi{10.1093/mnras/staa679}

\bibitem[{J.~Y. {Kim} {et~al.}(2018){Kim}, {Krichbaum}, {Lu}, {Ros}, {Bach}, {Bremer}, {de Vicente}, {Lindqvist}, \& {Zensus}}]{kim2018}
{Kim}, J.~Y., {Krichbaum}, T.~P., {Lu}, R.~S., {et~al.} 2018, \bibinfo{title}{{The limb-brightened jet of M87 down to the 7 Schwarzschild radii scale},} \aap, 616, A188, \dodoi{10.1051/0004-6361/201832921}

\bibitem[{Y. Kovalev {et~al.}(2007)Kovalev, Lister, Homan, \& Kellermann}]{kovalev2007}
Kovalev, Y., Lister, M., Homan, D., \& Kellermann, K. 2007, \bibinfo{title}{The Inner Jet of the Radio Galaxy M87,} The Astrophysical Journal Letters, 668, L27, \dodoi{10.1086/522603}

\bibitem[{R.-S. {Lu} {et~al.}(2014){Lu}, {Broderick}, {Baron}, {Monnier}, {Fish}, {Doeleman}, \& {Pankratius}}]{lu2014}
{Lu}, R.-S., {Broderick}, A.~E., {Baron}, F., {et~al.} 2014, \bibinfo{title}{{Imaging the Supermassive Black Hole Shadow and Jet Base of M87 with the Event Horizon Telescope},} \apj, 788, 120, \dodoi{10.1088/0004-637X/788/2/120}

\bibitem[{R.-S. {Lu} {et~al.}(2023){Lu}, {Asada}, {Krichbaum}, {Park}, {Tazaki}, {Pu}, {Nakamura}, {Lobanov}, {Hada}, {Akiyama}, {Kim}, {Marti-Vidal}, {G{\'o}mez}, {Kawashima}, {Yuan}, {Ros}, {Alef}, {Britzen}, {Bremer}, {Broderick}, {Doi}, {Giovannini}, {Giroletti}, {Ho}, {Honma}, {Hughes}, {Inoue}, {Jiang}, {Kino}, {Koyama}, {Lindqvist}, {Liu}, {Marscher}, {Matsushita}, {Nagai}, {Rottmann}, {Savolainen}, {Schuster}, {Shen}, {de Vicente}, {Walker}, {Yang}, {Zensus}, {Algaba}, {Allardi}, {Bach}, {Berthold}, {Bintley}, {Byun}, {Casadio}, {Chang}, {Chang}, {Chang}, {Chen}, {Chen}, {Chilson}, {Chuter}, {Conway}, {Crew}, {Dempsey}, {Dornbusch}, {Faber}, {Friberg}, {Garc{\'\i}a}, {Garrido}, {Han}, {Han}, {Hasegawa}, {Herrero-Illana}, {Huang}, {Huang}, {Impellizzeri}, {Jiang}, {Jinchi}, {Jung}, {Kallunki}, {Kirves}, {Kimura}, {Koay}, {Koch}, {Kramer}, {Kraus}, {Kubo}, {Kuo}, {Li}, {Lin}, {Liu}, {Liu}, {Lo}, {Lu}, {MacDonald}, {Martin-Cocher}, {Messias}, {Meyer-Zhao}, {Minter}, {Nair}, {Nishioka}, {Norton},
  {Nystrom}, {Ogawa}, {Oshiro}, {Patel}, {Pen}, {Pidopryhora}, {Pradel}, {Raffin}, {Rao}, {Ruiz}, {Sanchez}, {Shaw}, {Snow}, {Sridharan}, {Srinivasan}, {Tercero}, {Torne}, {Traianou}, {Wagner}, {Walther}, {Wei}, {Yang}, \& {Yu}}]{lu2023}
{Lu}, R.-S., {Asada}, K., {Krichbaum}, T.~P., {et~al.} 2023, \bibinfo{title}{{A ring-like accretion structure in M87 connecting its black hole and jet},} \nat, 616, 686, \dodoi{10.1038/s41586-023-05843-w}

\bibitem[{J.~C. McKinney \& C.~F. Gammie(2004)McKinney \& Gammie}]{mckinney2004measurement}
McKinney, J.~C., \& Gammie, C.~F. 2004, \bibinfo{title}{A measurement of the electromagnetic luminosity of a Kerr black hole,} The astrophysical journal, 611, 977

\bibitem[{J.~C. McKinney \& R. Narayan(2007)McKinney \& Narayan}]{mckinney2007disc}
McKinney, J.~C., \& Narayan, R. 2007, \bibinfo{title}{Disc--jet coupling in black hole accretion systems--I. General relativistic magnetohydrodynamical models,} Monthly Notices of the Royal Astronomical Society, 375, 513

\bibitem[{F. {Mertens} {et~al.}(2016){Mertens}, {Lobanov}, {Walker}, \& {Hardee}}]{mer2016}
{Mertens}, F., {Lobanov}, A.~P., {Walker}, R.~C., \& {Hardee}, P.~E. 2016, \bibinfo{title}{{Kinematics of the jet in M 87 on scales of 100-1000 Schwarzschild radii},} \aap, 595, A54, \dodoi{10.1051/0004-6361/201628829}

\bibitem[{M. {Mo{\'s}cibrodzka} {et~al.}(2017){Mo{\'s}cibrodzka}, {Dexter}, {Davelaar}, \& {Falcke}}]{monika2017}
{Mo{\'s}cibrodzka}, M., {Dexter}, J., {Davelaar}, J., \& {Falcke}, H. 2017, \bibinfo{title}{{Faraday rotation in GRMHD simulations of the jet launching zone of M87},} \mnras, 468, 2214, \dodoi{10.1093/mnras/stx587}

\bibitem[{M. {Nakamura} {et~al.}(2018){Nakamura}, {Asada}, {Hada}, {Pu}, {Noble}, {Tseng}, {Toma}, {Kino}, {Nagai}, {Takahashi}, {Algaba}, {Orienti}, {Akiyama}, {Doi}, {Giovannini}, {Giroletti}, {Honma}, {Koyama}, {Lico}, {Niinuma}, \& {Tazaki}}]{nakamura2018}
{Nakamura}, M., {Asada}, K., {Hada}, K., {et~al.} 2018, \bibinfo{title}{{Parabolic Jets from the Spinning Black Hole in M87},} \apj, 868, 146, \dodoi{10.3847/1538-4357/aaeb2d}

\bibitem[{R. Narayan {et~al.}(2007)Narayan, McKinney, \& Farmer}]{narayan2007}
Narayan, R., McKinney, J., \& Farmer, A. 2007, \bibinfo{title}{Self‐similar force‐free wind from an accretion disc,} Monthly Notices of the Royal Astronomical Society, 375, 548 , \dodoi{10.1111/j.1365-2966.2006.11272.x}

\bibitem[{A. Pandya {et~al.}(2016)Pandya, Zhang, Chandra, \& Gammie}]{pandya2016polarized}
Pandya, A., Zhang, Z., Chandra, M., \& Gammie, C.~F. 2016, \bibinfo{title}{Polarized synchrotron emissivities and absorptivities for relativistic thermal, power-law, and kappa distribution functions,} The Astrophysical Journal, 822, 34

\bibitem[{J. {Park} {et~al.}(2019){Park}, {Hada}, {Kino}, {Nakamura}, {Hodgson}, {Ro}, {Cui}, {Asada}, {Algaba}, {Sawada-Satoh}, {Lee}, {Cho}, {Shen}, {Jiang}, {Trippe}, {Niinuma}, {Sohn}, {Jung}, {Zhao}, {Wajima}, {Tazaki}, {Honma}, {An}, {Akiyama}, {Byun}, {Kim}, {Zhang}, {Cheng}, {Kobayashi}, {Shibata}, {Lee}, {Roh}, {Oh}, {Yeom}, {Jung}, {Oh}, {Kim}, {Hwang}, \& {Hagiwara}}]{park2019}
{Park}, J., {Hada}, K., {Kino}, M., {et~al.} 2019, \bibinfo{title}{{Kinematics of the M87 Jet in the Collimation Zone: Gradual Acceleration and Velocity Stratification},} \apj, 887, 147, \dodoi{10.3847/1538-4357/ab5584}

\bibitem[{T. {Piran}(2004){Piran}}]{tsvi04}
{Piran}, T. 2004, \bibinfo{title}{{The physics of gamma-ray bursts},} Reviews of Modern Physics, 76, 1143, \dodoi{10.1103/RevModPhys.76.1143}

\bibitem[{H.-Y. Pu {et~al.}(2022)Pu, Asada, \& Nakamura}]{pu2022modeling}
Pu, H.-Y., Asada, K., \& Nakamura, M. 2022, \bibinfo{title}{Modeling Nearby Low-Luminosity Active-Galactic-Nucleus Jet Images at All VLBI Scales,} Galaxies, 10, 104

\bibitem[{H.-Y. {Pu} \& M. {Takahashi}(2020){Pu} \& {Takahashi}}]{pu2020}
{Pu}, H.-Y., \& {Takahashi}, M. 2020, \bibinfo{title}{{Properties of Trans-fast Magnetosonic Jets in Black Hole Magnetospheres},} \apj, 892, 37, \dodoi{10.3847/1538-4357/ab77ab}

\bibitem[{H.-Y. Pu {et~al.}(2016)Pu, Yun, Younsi, \& Yoon}]{pu2016odyssey}
Pu, H.-Y., Yun, K., Younsi, Z., \& Yoon, S.-J. 2016, \bibinfo{title}{Odyssey: a public GPU-based code for general relativistic radiative transfer in Kerr spacetime,} The Astrophysical Journal, 820, 105

\bibitem[{H. Ro {et~al.}(2023)Ro, Kino, Sohn, Hada, Park, Nakamura, Cui, Yi, Chung, Hodgson, {et~al.}}]{ro2023spectral}
Ro, H., Kino, M., Sohn, B.~W., {et~al.} 2023, \bibinfo{title}{Spectral analysis of a parsec-scale jet in M 87: Observational constraint on the magnetic field strengths in the jet,} Astronomy \& Astrophysics, 673, A159

\bibitem[{K. Takahashi {et~al.}(2018)Takahashi, Toma, Kino, Nakamura, \& Hada}]{takahashi2018fast}
Takahashi, K., Toma, K., Kino, M., Nakamura, M., \& Hada, K. 2018, \bibinfo{title}{Fast-spinning black holes inferred from symmetrically limb-brightened radio jets,} The Astrophysical Journal, 868, 82

\bibitem[{A. Tchekhovskoy {et~al.}(2008)Tchekhovskoy, McKinney, \& Narayan}]{tchekhovskoy2008simulations}
Tchekhovskoy, A., McKinney, J.~C., \& Narayan, R. 2008, \bibinfo{title}{Simulations of ultrarelativistic magnetodynamic jets from gamma-ray burst engines,} Monthly Notices of the Royal Astronomical Society, 388, 551

\bibitem[{R.~C. Walker {et~al.}(2018)Walker, Hardee, Davies, Ly, \& Junor}]{walker2018structure}
Walker, R.~C., Hardee, P.~E., Davies, F.~B., Ly, C., \& Junor, W. 2018, \bibinfo{title}{The structure and dynamics of the subparsec jet in M87 based on 50 VLBA observations over 17 years at 43 GHz,} The Astrophysical Journal, 855, 128

\bibitem[{R.~C. {Walker} {et~al.}(2008){Walker}, {Ly}, {Junor}, \& {Hardee}}]{walker2008_movie}
{Walker}, R.~C., {Ly}, C., {Junor}, W., \& {Hardee}, P.~J. 2008, in Journal of Physics Conference Series, Vol. 131, Journal of Physics Conference Series (IOP), 012053, \dodoi{10.1088/1742-6596/131/1/012053}

\end{thebibliography}
\bibliographystyle{aasjournalv7}

\end{document}